\documentclass[article]{JHEP3}

\usepackage{epsfig,multicol}

\newcommand{\be}{\begin{equation}}
\newcommand{\ee}{\end{equation}}
\newcommand{\bea}{\begin{eqnarray}}
\newcommand{\eea}{\end{eqnarray}}
\newcommand{\SG}{\sigma}
\newcommand{\ep}{\varepsilon}
\newcommand{\nn}{\nonumber}

\title{ Minkowski solution of Dyson-Schwinger equations 
in momentum subtraction scheme}

\author{Vladim\'{\i}r \v{S}auli\\
Department of Theoretical Physics, 
Nuclear Physics Institute, \v{R}e\v{z} near Prague,CZ-25068\\
E-mail: \email{sauli@ujf.cas.cz}}

\abstract{
Using the Green's function integral representation the Dyson-Schwinger equations are
solved directly in Minkowski space.
Essential ideas of  spectral techniques 
are discussed  and applied on two renormalizable models: the Yukawa theory 
with massive pseudoscalar meson and conventional  spinor QED.
Within the momentum subtraction procedure, the  renormalization  is performed 
analytically which  leads to the usual dispersion formulation.
The  electron propagator obtained in this frame is  
compared with  the solution of Euclidean Dyson-Schwinger equation and with the 
perturbation theory results as well.
The  proposed method has the advantage of  obtaining   solutions 
in  both the space- and {\bf{time}}-like regimes of momenta. In addition, when the 
coupling constant increases we find some unexpected discrepancy
between the Euclidean and integral representation solutions.
Especially for the supercritical couplings, the propagator pole is absent 
and there is no solution for spectral fermion function.
}


\begin{document}

\section{Introduction}

There are considerable  interest in  studies of strong coupling quantum field
theories since many interesting phenomena, are believed have nonperturbative
origins. In renormalizable field theory  it is desirable to construct renormalization 
techniques which completely respect  the symmetries of the underlying theory. 
In comparison with  the conventional perturbation theory the nonperturbative treatment 
is more complicated and much effort is needed to perform renormalization procedure 
in a well-controlled manner. The fully regularization-independent method is proposed 
through the Dyson-Schwinger \cite{DYSON},\cite{SCHWINGER}  formalism, which is the main motivation of this study. 
The other motivation of the presented work is to avoid a constrains of Euclidean metric.
Therefore the main part of our calculation is performed directly in Minkowski space and we offer the 
obtained numerical solution of the Dyson-Schwinger equations (DSEs)  
at all  regimes of momenta. If the Euclidean analogue of solution is known and already present
in the literature we perform the appropriate calculation of the Euclidean DSEs 
and compare with Minkowski solution.

 Contemporary application of DSEs in hadronic physics 
offers rigorous insight into the infrared domain of Quantum Chromodynamics  
\cite{SMEKAL},\cite{ROBERTS},\cite{ROBERTS2} at zero temperature as well as 
DSEs provide solid foundation  
for nonperturbative continuous approach at nonzero temperature and density \cite{ROBERTO}.
Without the solution of DSEs it is  difficult   to reach any 
reasonable result on dynamical mass generation in the (Extended, Walking,..) Technicolor
models \cite{GEORGI},\cite{LOVE},\cite{KING}. In these models (usually without higgses)
the values of the various condensates and the masses of the Standard model particle content are obtained 
from the solution of the appropriate gap equations.  In practice, the system of the DSEs 
is truncated  by an approximation of the Green functions that were thrown away. 
In addition, the truncated set of DSEs must be renormalized and as it is usually required by the 
method, the loop integral must be regularized before proper renormalization step.
When any S-matrix element is  completed from the Green functions the definite results
must be 1.renormgroup invariant, 2.gauge invariant 3.gauge fixing independent.
All the three points are well understood in the   perturbation theory treatment and it is desirable 
to build the similar nonperturbative method. The Pinch Technique  
\cite{CORNWALL},\cite{JANDJ},\cite{BINOSI} and the Background-Field Method 
\cite{DENNER} should offer the gauge fixing  independent Green's functions, 
however how to fulfill all three aforementioned points  is not answered by  fully satisfactory  way
in DSEs treatment. In this paper, we particularly concern on the points 1 and 2 mentioned above.

It is clear that the truncation of the DSEs system, if it is improperly performed,  
can violate gauge identity of the underlying gauge theory. In an Abelian versions 
of gauge field theory (scalar and spinor quantum electrodynamic) this problem was 
already solved by at least  two technically different ways. Ball and Chiu  \cite{BALLCHIU}
have derived the formula for the QED (truncated, proper) vertex that allows to 
close the  DSEs system by a unique gauge covariant way (the lowest point Green's functions 
satisfy Ward-Takahashi identities). The second successful approach is the longstanding 
method  known as the 'Gauge Technique' \cite{SALAM},\cite{CIMENTO},
\cite{DELBOURGO}, \cite{HOSHINO}. In this treatment the gauge covariant ansatze for the full 
(untruncated, i.e., not proper) vertex function is written in a terms of matter field 
spectral function. The technical advantage of the Gauge Technique is that the resultant
equation for fermion propagator is linearized in spectral function. 
Due to this simplification, the Gauge Technique  offers a solutions in compact analytical form. 
Of course, 
this linearization does not take place in the equation for the photon propagator and
in fact this is one of the approach essential weakness: it cannot be true in general 
even for the case of electron propagator. Anticipate here, that no linearization does appear
in this presented work and the appropriate integral equations are solved in their full form
(they are third and second order integral equations). In the other side, the aforementioned linearization 
of Gauge Technique obtained equations do not exclude their reliability in the soft coupling regime   
and we shall mention the  Gauge Technique once again when we discus infrared limit of 
electron propagator. 

In the last decade  number of papers dealing with some improved vertices in  QED DSEs studied
the connection to dynamical mass generation \cite{PENNINGTON},\cite{CURPEN},\cite{BLOCH},
or they have been subjected to the various renormalization
scheme used in its nonperturbative context \cite{HAWE},\cite{SIZE},\cite{WILL2},\cite{WILDA} 
(for a review of the earlier 
attempts see \cite{ROBERTS}). In the paper \cite{WILDA}, the modified cut-off regularization method has been compared with the nonperturbative dimensional regularization scheme. This study exemplified the nonperturbative 
equivalence of different regularization-renormalization schemes. It also faces how  careful
the appropriate numerical procedure must be in order to obtain the reliable physical results 
(without breaking gauge and Poincar\'e invariance of the underlying theory). 
 We extend this studies 
to the case of (direct) momentum subtraction  (MOM) scheme where neither numerical regularization 
is used. Likewise in the perturbation treatment,  the all appropriate loop integrals are subtracted at  certain value of 
external momenta and being then finite they are integrated analytically. 
Such a procedure leads to the usual dispersion relations for renormalized proper functions.
We use the   Green's function spectral representation which allows to convert 
momentum DSEs to the real equations for the spectral functions. At this point this 
approach is similar to the treatment used in the works \cite{WUZHANG},\cite{IMAG} and of course,
to the one used in  
Gauge Technique studies. Note at this place, that  very similar method 
already has succeeded in the case of simple scalar  models \cite{JAABSE}, \cite{SAULI},\cite{META}
where the on-shell renormalization scheme was pronounced. 

It is worthwhile meaning that making some approximation the regularization independent 
and analytical answer can be obtained \cite{WILDA},\cite{KIZIL},\cite{ATKINSON}. 
In very  early stage of DSEs study there was made a sophisticated study \cite{EDWARDS} of spinor QED
where the propagators entering  calculation are taken as bare ones and  the resulting vertex turns out 
to be a hypergeometric function in that case. Furthermore, in rather special case 
(for massless fermion in rainbow and quenched Feynman-Fermi gauged QED)
the authors of \cite{LADDER} found results analytically.

In our paper the method of solution will be illustrated at  two  model cases.
The first of them is the Yukawa Theory (YT), i.e., the theory of fermions interacting with
spinless pseudoscalar boson. 
 The Yukawa  Lagrangian reads explicitly:  
\bea
{\cal L}&=&i\bar{\Psi}\not\!\partial\Psi-m_{0}\bar{\Psi}\Psi+
\frac{1}{2}\partial_{\mu}\Phi \partial^{\mu}\Phi
-\frac{m_{0\phi}^2}{2}\Phi^{2}
\nn \\
&-&g_{0}i\bar{\Psi}\gamma_{5}\Psi\Phi-h_0\Phi^4/, ,
\eea
 where $m_{0} $ , $m_{0\phi}$ are bare masses of fermion and meson 
corresponding to the unrenormalized fields $\Psi_f$ and $\Phi$ respectively
 and $g_0$ , $h_0 $ represent  the unrenormalized values of coupling constants.
For the sake of simplicity, the Yukawa vertex is modeled by its tree value.
In  addition, it is assumed that $h<<g$,  therefore we also neglect the quartic 
mesons self-interaction. Remind here, that the quartic term in $\Phi$ is necessary due to the 
general requirement of renormalizability.  The counter-term part has to cancel infinities
appearing in the fermion loops contribution to scattering 
process $\Phi\Phi\rightarrow \Phi\Phi$. Here, the contribution from such  process does not enter 
our DSEs due to their truncation mentioned above and an omission of $h_0\Phi^4 $ represents 
self-consistent approximation.    

The second model we employ is a conventional spinor QED.
The  approach is discussed for a general case of gauge covariant vertex.
The numerical results are presented  for the quenched, rainbow approximation 
leaving the complete treatment for the later presentation.
Approximation used in this article   has the advantage of being 
simple enough, which makes it an excellent testing ground for the proposed  Minkowski analysis.  
The so-called ladder approximation (the bare vertex is used, the name follows from the ladder 
approximation of the Bethe-Salpeter equation) is generally believed to be reliable in the Landau 
gauge only  $\xi=0$, and therefore we have used that gauge. Having all the numerical solutions stable
and making comparison between Euclidean and Minkowski results we found
that they agree only when the coupling is small enough. Increasing the coupling constant
the obvious  discrepancy does appear. Approaching QED coupling $\alpha=e^2/(4\pi)$
to its critical value $\alpha=\pi/3$ we are leaving only with Euclidean solution, whilst the
spectral Minkowski equation tends to flaw.     

The necessary analytical formulas  are reviewed in the next Section. It involves dispersion 
relation technique, its connection with MOM and derivation of unitary equations for  DSEs.
The Section III is particularly devoted to the QED, where also the Euclidean version of fermion DSE
is reviewed. Numerical results are presented in the following Section IV.

Some details of the calculation and of the numerical method are explained in the Appendices.

\section{Spectral representation, Analyticity and Renormalization}

In the following sections we give  overview of  some basic facts about the 
Green's function Spectral Representation (SR), 
Dispersion Relation (DR) technique  and their relation to the renormalization.
 
 The Lehmann  representation \cite{Lehmann},\cite{KALLEN} for propagator
 can be derived from  Lorentz covariance and  quantum
mechanical requirement on the positivity of  spectral density. 
The relations  (\ref{s-fermici}) display the necessary 
SRs for appropriate propagators entering the calculation here:
\bea         \label{s-fermici}
G(p^2)&=&\int d\omega \frac{\bar{\SG}(\omega)}{p^2-\omega+i\epsilon}
=\frac{r_{\phi}}{p^2-m_{\phi}^2+i\epsilon}+\int d\omega
\frac{\SG(\omega)}{p^2-\omega+i\epsilon}
\nn \\
S_f(p)&=&\int d\omega \frac{\not\!p\bar{\SG}_v(\omega)+\bar{\SG}_s(\omega)}
{p^2-\omega+i\epsilon}
=\frac{r_f}{\not\!p-m}+
\int d\omega
\frac{\not\!p\SG_v(\omega)+\SG_s(\omega)}{p^2-\omega+i\epsilon}
\nn \\
G^{\mu\nu}(p)&=&
\left(-g^{\mu\nu}+\frac{p^{\mu}p^{\nu}}{p^2}\right)G_{T}(p^2)
-\xi \frac{p^{\mu}p^{\nu}}{p^2}
\nn \\
G_{T}(p^2)&=&\int d\omega \frac{\bar{\SG}_{\gamma}(\omega)}{p^2-\omega+i\epsilon}
=\frac{r_{\gamma}}{p^2+i\epsilon}+\int d\omega
\frac{\SG_{\gamma}(\omega)}{p^2-\omega+i\epsilon}.
\eea
Here $G,S_f $ are the full propagators of 
particles with spin $ 0$ and $\frac{1}{2}$  respectively.
$G^{\mu\nu}$ describes propagation of massless particle which corresponds to the
gauge field. The longitudinal $\xi$-dependent part follows 
from the covariant linear gauge fixing $g_A=\delta(\xi-\partial_{\mu}A^{\mu})$ of quantum action.
The single particle contributions $r_i\delta(\omega-m_i^2)$ are integrated out 
from  the  full spectrum $\bar{\SG}$ and it is  assumed 
that remaining weight functions $\SG$'s in (\ref{s-fermici}) are smoothed 
ones and not complicated distributions at all.

It is notable in this place, that there is less formal proof of SRs for the simple Quantum field models.
These models are represented by   the scalar quantum field theories without 
derivative interaction.
The existence of integral representation was proved to the all orders of perturbation theory  even for an arbitrary $n-$points
Green's function. In this case, the so called Perturbation Theory Integral Representation (PTIR) 
was derived by Nakanishi \cite{NAKAN}. Furthermore, it was shown that the PTIR is unique, which property
appears to be very useful in some applications (see for instance \cite{JAABSE},\cite{META}, and references herein).

Although the  formal derivation of Lehmann representation is  rather straightforward  
(see some standard textbooks \cite{ZUBER}, \cite{SCHWEBER}), 
 it necessarily breaks down when considering a theory without free particle asymptotic states, 
 i.e. the theory with confinement. 
If the confinement takes place in a given theory then the particle can never be on mass shell and the 
appropriate propagator should not posses  singularities at the real time-like axis of $p^2$. 
The absence of Lehmann representation  should  be a good signal for confinement
\cite{HONZA},\cite{PETA}. Actually, as we will show in the Section devoted to the strong coupling 
quenched QED, the disappearance of solution for Lehmann function rather sharply coincides with the transition
to a confined phase.

As was mentioned,  the nonperturbative 
regularization - renormalization  procedure is not so transparent as  
in the case of  perturbative treatment. Our  task is  whether the renormalization procedure should be performed analytically  
in  an easy and transparent fashion as it is in the perturbative approach.  
First of all we must answer the question that  naturally arises: 
what are the physical criteria, which 
will one to determined the solution of DSEs as a physically meaningful. 
In particular, the following is required for two point function:

\

(1) The renormalized  full Green's function 
satisfies its own renormgroup equation
$\gamma=\mu\frac{d}{d\mu}G(p^2,\mu)$, where gamma represents logarithmic differentiation of conventionally defined
field strength renormalization constant $Z$
(explicitly introduced later by Eqs. (\ref{cauparde}))  $\gamma=-\mu\frac{dZ}{d\mu},$
i.e., the unrenormalized propagators $S_{f0},G_{0},G^{\mu\nu}_0$
have to be manifestly independent on the choice of renormalization scale.  

\

(2) The on-shell renormalization scheme (ORS) should be a
 special choice of the renormalization scale $\mu=m$.
Furthermore, the  position of the pole $p^2=m^2$ is renormgroup invariant quantity. 

\

(3) The  Green's functions  are an analytical ones. 
Up to a  subtracting polynom, the real and imaginary parts of proper 
Green's function (one particle irreducible diagram with truncated legs)
is uniquely defined by a dispersion relation.

\

(4) The off-shell proper Green's functions calculated within DSE approach 
should admit renormalization.   It implies that the possible maximum 
number of subtraction is two for self-energy  and at most one 
for the possible triplet and quartic vertex renormalization. All the coefficients in the 
appropriate subtracting polynom
should be absorb-able in the counter-term part of the original Lagrangian.

\

(5) As far as it is possible, the renormalization procedure should respect  
classical symmetry of the theory. Particularly, when dealing with gauge theory, 
the  Green's functions should satisfy Ward-Takahashi identities.

 In the next subsections  we review the general framework of analytical  renormalization technique
and explain the renormalization scheme which  is  actually  demonstrating
that the Green's function obtained by this technique
satisfy all the above requirement

\subsection{ Renormalization of DSE for Yukawa pseudoscalar }
%SSSSSSSSSSSSSSSSSSSSSSSSSSSSSSSSSSSSSSSSSSSSSSSSSSSSSSSSSSSSSSSSSSSSSSSSSSSSSSSSSSSSSSSSSSSSSSSSSSSSSSSSSSSS

Here we start with the discussion 
of  boson propagator.
 The unrenormalized version of DSE reads

\bea  \label{nonren}
G_0^{-1}(p^2)&=& p^2-m_0^2-\widetilde{\Pi}_0(p^2)
\nn \\
\widetilde{\Pi}_0(p^2)&=&ig_0^2\int Tr \frac{d^4l}{(2\pi)^4}
\Gamma_{Y0}(p,l)S_{0}(l)\gamma^{5}S_{0}(l-p),
\eea 
where the index $0$  represents bare quantities.
To renormalize  YT we conventionally introduce the   renormalization 
functions with the appropriate counter-terms. 
Choosing  some arbitrary  renormalization scale $\mu^2$ they are:
\bea \label{cauparde}
\Phi=Z_{\phi}^{1/2}\phi_R &\quad,&
\Psi=Z_{\Psi}^{1/2}\Psi_R
\nn \\ 
m_{\phi 0}=Z_{m_{\phi}} m_{\phi}(\mu); \quad \delta_{m_{\phi}}=m_0^2-m(\mu)^2 &\quad,&
m_0=Z_{m} m(\mu); \quad \delta_{m}=m_0-m(\mu) 
\nn \\
g_0=\frac{Z_g^{1/2}}{Z_{\Psi}Z_{\phi}^{1/2}}g &\quad,&
h_0=\frac{Z_h^{1/2}}{Z_{\phi}^2}h \quad ,
\eea
where $Z_{\Phi}$ ($Z_{\Psi}$)is a renormalization  boson (fermion) field-strength constant , $ Z_{m} $
is the mass renormalization constant, $m_0 $ is a bare mass while $m(\mu)$ represents renormalized mass.
Furthermore, $Z_h$ and $Z_g$ represent the renormalization constant of  quartic 
$\Phi^4$ and triplet $\bar{\Psi}\Psi\Phi$ vertex respectively, i.e. for instance
 we can  write for the renormalized proper triplet vertex:$\Gamma_{Y}(p,p-k)=Z_g\Gamma_{Y0}(p,p-k)$.
First of all we renormalize the propagators with respect to the field renormalization only i.e.
$G_0=Z_{\phi}G$, $S_0=Z_{\psi}S $. For this purpose we multiply the unrenormalized DSE (\ref{nonren})
by the constant $Z_{\phi}$. A simple algebra gives
\bea   \label{posuk}
G^{-1}(p^2)&=& Z_{\phi}(p^2-m_0^2)-\Pi_0(p^2)
\nn \\
\Pi_0(p^2)&=&ig^2\int Tr \frac{d^4l}{(2\pi)^4}
\Gamma_Y(p,l)S_f(l)\gamma^{5}S_f(l-p) \quad ,
\eea 
where the  proper vertex $\Gamma_Y$  satisfies its own DSE.
To truncate the system of DSE we make most simple approximation
 $\Gamma_Y(p,l)=\gamma^{5}$.
Since the pseudoscalar particle $\Phi$ requires quadratically divergent mass renormalization and 
logarithmically divergent field strength renormalization the relation between renormalized and
unrenormalized self-energy function must be of the form
\bea \label{disperze1}
\Pi_0(p^2)&=&\Pi_0(\mu^2)+ 
\Pi_0'(\mu^2)(p^2-\mu^2)
+\Pi(\mu;p^2)\, ,
\eea
where the renormalized self-energy satisfies double subtracted DR
\be  \label{cojeco}
\Pi(\mu;p^2)=\int d\omega\frac{\rho_\phi(\omega)}{(p^2-\omega+i\epsilon)}
\left[\frac{p^2-\mu^2}{\omega-\mu^2}\right]^2\, .
\ee
Of course, from the Rel. (\ref{disperze1}) we obtain the standard receipt for calculation 
of the propagator in  MOM   scheme  
\bea\label{subtrakce}
G^{MOM}(\mu;p^2)&=&\left(p^2-m^2(\mu)-\Pi(\mu;p^2)\right)^{-1}
\nn \\
\Pi(\mu;p^2)&=& \Pi_0(p^2)- \Pi_0(\mu^2)
-\frac{d}{d p^2}\Pi_0(p^2)|_{p^2=\mu^2}(p^2-\mu^2)\, .
\eea     
 From Eq.(\ref{disperze1}), one can readily see that the unrenormalized 
value $\Pi_0(\mu^2)$ corresponds to the dominant part of the mass renormalization 
constant and the derivative $\Pi_0'(\mu^2)$ corresponds to 
the field-strength renormalization, explicitly we have 
\bea
Z_{\phi}&=&1+\Pi_0'(\mu)
\nn \\
Z_{m_{\phi}}^2&=&Z_{\phi}^{-1}\left(1-\frac{\Pi_0(\mu^2)-\Pi_0'(\mu^2)\mu^2}{m^2(\mu)}\right)
\eea

Note that in the perturbative context this  schemes sometimes referred as the BPHZ renormalization scheme
 and   the appropriate subtraction procedure
 is  called (Bogoliubov) R-operation. 
In the next text  we drop out
the labeling MOM since no other renormalization scheme is used throughout  this article.

Employing the well known 
functional identity for distributions
\be
\frac{1}{x'-x+i\ep}=P. \frac{1}{x'-x}-i\pi\delta(x'-x)
\ee
we see that the function $\pi\rho(p^2)$ represents the absorptive (imaginary) 
part of renormalized self-energy $\Pi(p^2)$ as well as of the unrenormalized one.
The last  statement is unrelated with the question of (in)finiteness of the counter-terms, 
since such  renormalization procedure has nothing directly 
with the presence of infinities 
and can be consistently applied to a theories which are UV convergent. The detailed derivation 
of the weight function $\rho_{\phi}$ is relegated to the Appendix A. Here we simply review the result:
\bea \label{substrakce}
&&\rho_{\phi}(\omega)= \left(\frac{g}{2\pi}\right)^2
\left[r_f^2\frac{\omega}{2}\sqrt{1-\frac{4m^2}{\omega}}\Theta(\omega-4m^2)\right.
\nn \\
&&+mr_f\int\limits_{(m+m_{\phi})^2}^{\infty}d\beta
\left(\frac{1}{2}(\omega-m^2-\beta)\SG_v(\beta)+m\SG_s(\beta)\right)
X(\omega;m^2,\beta)
\nn \\
&&+\left.\int\limits_{(m+m_{\phi})^2}^{\infty}d\alpha \int\limits_{(m+m_{\phi})^2}^{\infty}d\beta 
\left(\frac{1}{2}(\omega-\alpha-\beta)\SG_v(\alpha)\SG_v(\beta)+
\SG_s(\alpha)\SG_s(\beta)\right)
X(\omega;\alpha,\beta)\right]
\eea    
where  the function $X$ can be expressed through the triangle function $\lambda$ by the following way
\be 
X(a;b,c)=\frac{\lambda^{1/2}(a;b,c)}{a}\Theta(a-(\sqrt(b)+\sqrt(c))^2)
\ee
and $m$ is a pole mass of fermion.

A boson pole mass $m_{\phi}=m_{\phi}(m{\phi})$  is conventionally defined as
$G^{-1}(m^2_{\phi})=0$. From its definition it reads:
\be  \label{klokan}
m_{\phi}^2(\mu)=m_{\phi}^2+\int d\omega\frac{\rho_\phi(\omega)}{\omega-m_{\phi}^2}
\left[\frac{m_{\phi}^2-\mu^2}{\omega-\mu^2}\right]^2.
\ee

Due to the algebraic simplicity of the DRs in  MOM scheme we can 
immediately recognize that the inverse propagators $G$ renormalized at two different scales $\mu,\mu'$
just differ by finite polynom $a_{fin.}+b_{fin}p^2$. Written explicitly, the relation reads 
\bea
G^{-1}(\mu,p^2)&=&a_{fin.}+b_{fin}p^2+G^{-1}(\mu',p^2)
\nn \\
a_{fin.}&=&\frac{\Pi(\mu;\mu'^2)+\Pi(\mu';\mu^2)}{\mu^2-\mu'^2}= 
\int d\omega\frac{\rho_\phi(\omega)(\mu'^2-\mu^2)^2}
{(\omega-\mu'^2)(\omega-\mu^2)^2}
\nn \\
b_{fin.}&=&m^2(\mu')-m^2(\mu)+\frac{\mu'^2\Pi(\mu';\mu^2)+\mu^2\Pi(\mu;\mu'^2)}{\mu'^2-\mu^2}
\nn \\
&=& m^2(\mu')-m^2(\mu)+
\int d\omega\frac{\rho_\phi(\omega)\omega(\mu'^2-\mu^2)}
{(\omega-\mu'^2)(\omega-\mu^2)^2}.
\eea
The MOM identities (\ref{laco}) exhibit the evolution of the propagator within the change of the renormalization scale.

\subsection{Renormalization of DSE for Yukawa fermion}
%FFFFFFFFFFFFFFFFFFFFFFFFFFFFFFFFFFFFFFFFFFFFFFFFFFFFFFFFFFFFFFFFFFFFFFFFFFFFFFFFFFFFFFFFFFFFFF

The extension to the fermion case  proceeds   similar way.
 DSE for fermion propagator reads
\be \label{fermprop}
    S_f^{-1}(\not\!p) 
         =  Z_{\Psi} [\not\!p - m_0]
                              - \Sigma_0(p), 
\ee
where $m_0$ represents  bare mass of fermion and the renormalization constant $Z_{\psi}$
was already introduced in (\ref{cauparde}). It is convenient to split
unrenormalized self-energy to its dirac vector and dirac scalar part
\be    
\Sigma_0(p)=\not\!pa_0(p^2)+b_0(p^2)\,.
\ee
The scalar functions $a_0,b_0$ can be easily identified from the explicit expression for fermion self-energy
\be\label{zuzu}
\Sigma_0(\not\!p)=  
-ig^2\int\frac{d^4l}{(2\pi)^4}\Gamma_Y(p,l)
S_f(l)\gamma^{5}G((p-l)^2)\,.
\ee
The fermion self-energy is only logarithmically divergent
and one subtraction is sufficient to make the  scalar functions  
$a_0,b_0$ finite. The renormalized self-energy then reads

\be    
\Sigma(\mu,p)=\not\!p a(\mu,p^2)+b(\mu,p^2),
\ee
where the subtraction leads to the  following DR's for renormalized $a,b$ 

\bea    \label{abdisperze}
a(\mu,p^2)&=& a_0(p^2)- a_0(\mu^2)=\int ds 
\frac{\rho_v(s)(p^2-\mu^2)}{(p^2-s+i\ep)(s-\mu^2)}
\nn \\
b(\mu,p^2)&=& b_0(p^2)- b_0(\mu^2)=\int ds 
\frac{\rho_s(s)(p^2-\mu^2)}{(p^2-s+i\ep)(s-\mu^2)} .
\nn \\
\eea 
Then the  renormalized version of  (\ref{fermprop}) can  be written  like

\be   \label{fermren}
 S_f^{-1}(p) 
   =  A(\mu)[\not\!p - m(\mu)]
   - \Sigma(\mu,p)  \quad .
\ee
where $m(\mu),(A(\mu))$ represents renormalized fermion mass (coefficient of $\not p$) fixed at the scale $\mu$.
As it is usually in DSE treatment the relation (\ref{fermren})
can be  equivalently rewritten into the form:
\bea    \label{zvyk}
S^{-1}(p)  &=&   A(p^2) \not\!p - B(p^2)\, ,
\nn \\
A(p^2)&\equiv& A(\mu)-a(\mu,p^2)\, ,
\nn \\
B(p^2)&\equiv& A(\mu)m(\mu)+b(\mu,p^2)\, ,
\eea
where  we do not indicate explicit dependence on $\mu$ in the renormalized functions $S,A,B$
for brevity. 
For the  absorptive parts $\Im  A=\Im  a=\pi \rho_v; \Im  B=\Im  b=\pi \rho_s$
we can find the following results:
\bea   \label{zadny}
\rho_v(\omega)&=& \frac{-g^2}{(4\pi)^2}
\left[ r_fr_{\phi}X_1(\omega;{m_{\phi}}^2,m^2)
+mr_{\phi}\int\limits_{(m+m_{\phi})^2}^{\infty}d\alpha
\SG_v(\alpha)X_1(\omega;m_{\phi}^2,\alpha)\right.
\nn \\
&+&r_{f}\left.\int\limits_{4m^2}^{\infty} d\beta
\rho_{\phi}(\beta)X_1(\omega;\beta,m^2)
+\int\limits_{4m^2}^{\infty} d\beta\int\limits_{(m+m_{\phi})^2}^{\infty}d\alpha
\rho_{\phi}(\beta)\SG_v(\alpha)X_1(\omega;\beta,\alpha)\right].
\eea
\bea \label{becko}
\rho_s(\omega)&=& \frac{g^2}{(4\pi)^2}
\left[r_{\phi}r_f mX(\omega;m_{\phi}^2,m^2)
+r_{\phi}\int\limits_{(m+m_{\phi})^2}^{\infty}d\alpha
\SG_s(\alpha)X(\omega;m_{\phi}^2,\alpha)\right.\, ,
\nn \\
&+& \left.\int\limits_{4m^2}^{\infty} d\beta
\rho_{\phi}(\beta)X(\omega;\beta,m^2)
+\int\limits_{4m^2}^{\infty} d\beta\int\limits_{(m+m_{\phi})^2}^{\infty}d\alpha
\rho_{\phi}(\beta)\SG_s(\alpha)
X(\omega;\alpha,\beta)\right]\, ,
\eea
where the function $X$ has been introduced earlier and 
$X_1(x;y,z)=\lambda^{1/2}(x;y,z)(x-y+z)/(2x^2)\Theta(x-(\sqrt(y)+\sqrt(z))^2)$. 
The  detailed derivation of Eq.'s (\ref{zadny}),(\ref{becko})is presented in the Appendix A.

The renormgroup invariant 
 mass function  which is conventionally defined as
\be  \label{dyn}
M(p^2)=B(p^2)/A(p^2).
\ee
Since the on shell mass $m\equiv m(m)$ is 
usually the one which is the best known experimentally,  it is useful to introduce the relation 
between $m$ and the one renormalized at $\mu$. From its definition $S(m)^{-1}=0$  we have
\bea \label{defonshell}
m&=&A(\mu)^{-1}[m(\mu)+a(\mu;m^2)m+b(\mu;m^2)].
\eea
We explicitly choose
\be \label{rencon}
A(\mu)=1
\ee
in this paper and Yukawa fermion  mass  $m(\mu)$ is fixed through (\ref{defonshell}) such that $m=1$.

Imposing the renormalization condition on $S$ at two different scales $\mu,$ and $\mu'$
we can again recognize  that the inverse of the fermion propagator 
differs by certain finite piece $c_{fin.}+\not pd_{fin.}$,
here we only note that the coefficients $c_{fin.},d_{fin.} $ can be expressed through
the functions $a,b(\mu;\mu'^2)$ and   $a,b(\mu';\mu^2)$.

%%%%%%%%%%%%%%%%%%%%%%%%%%%%%%%%%%%%%%%%%%%%%%%%%%%%%%%%%%%%%%%%%%%%%%%%%%%%%%%%%%%%%%%%%%%%%%%%%%%%%%%%%%%%%%%%%%%%%%

\subsection{ DSE for $\SG's$- The unitary equations }

 The derivation of  the DSE for Lehmann weights is presented in this section. 
 Evaluating the imaginary  
part of the trivial identity $1=G(p^2)G(p^2)^{-1}$ with
SR (\ref{s-fermici}) used for $G$ and  DR used for $\Pi$ in the DSE, i.e. in eq. 
 $G^{-1}=p^2-m_{\phi}(\mu)- \Pi(\mu;p^2)$ yields
\bea \label{pyjon}
\SG(\omega)(\omega-m^2_{\phi}(\mu))=r_{\phi}\frac{\rho_{\phi}(\omega)}{\omega-m_{\phi}^2}
+\{\rho_{\phi}  * \SG\},
\eea
where the symbol $\{\rho_{\phi}  * \SG\}$
represents following principal value integral:
\be \label{princip2}
P.\int dx\frac{\rho(s)\SG(x)+\SG(s)\rho(x)\frac{(s-\mu^2)^2}{(x-\mu^2)^2}}{s-x}.
\ee
The pseudoscalar propagator  residuum
\be
r_p=\lim_{p^2\rightarrow m_{\phi}^2}\frac{p^2-m_{\phi}^2}{G^{-1}(p^2)}
\ee
is most easily evaluated through the appropriate DR for self-energy
\be \label{rezion}
r_p=\frac{1}{1-\Pi'(m_{\phi}^2)}
=\left[1-\int d\omega\frac{\rho_\phi(\omega)(m_{\phi}^2-\mu^2)(m_{\phi}^2+\mu^2-2\omega)}
{(m_{\phi}^2-\omega)^2(\omega-\mu^2)^2}\right]^{-1}
\ee
It is not surprising that the equation (\ref{pyjon}) looks particularly simply in  
on mass-shell renormalization scheme:
\be \label{pyzon}
\SG(\omega)=\frac{\rho_{\phi}(\omega)}{(\omega-m_{\phi}^2)^2} 
-\frac{\{\rho_{\phi}  * \SG\}}{\omega-m_{\phi}^2} .
\ee
The fermion case can be treat by a very similar way. 
From the fermion  DSE we can obtain for the  residuum
\be \label{rezo}
r_f=\lim_{\not p \rightarrow m}\frac{\not p-m}{S^{-1}(p)}=
[1-a'(\mu;m^2)m-b'(\mu;m^2)-a(\mu;m^2)]^{-1}.
\ee
where we explicitly used the renormalization condition (\ref{rencon}).
Furthermore, writing the trivial identity $S^{-1}S-1=0$ in a suitable form:
\bea \label{lhss}
&&\left[\frac{r_f}{\not\!p-m}+\int d\omega \frac{\not\!p \SG_v(\omega)+m\SG_s(\omega)}
{p^2-\omega+i\ep}\right]\times
\nn \\
&&\left[\not\!p-m(\mu)-\int ds \frac{(\not\!p \rho_v(s)+\rho_s(s))(p^2-\mu^2)}{(p^2-s+i\ep)(s-\mu^2)}\right]=1
\eea
then we can easily arrived at the relations between Lehmann weights $\sigma$'s  
and the absorptive parts of self-energy $\rho$. Projecting the obtained result
by  $ \frac {Tr}{4\pi p^2} $ and $\frac{Tr}{4\pi mp^2} \not\!p $  leads to the coupled set
of integral equations
\bea  \label{Takk}
\SG_v(\omega)&=& \frac{f_1+m(\mu)f_2}{\omega-m^2(\mu)}
\quad ; \quad 
\SG_s(\omega)=\frac{m(\mu)f_1+\omega f_2}{\omega-m^2(\mu)}
\nn \\
f_1&\equiv&r\frac{\omega\rho_v(\omega)+m\rho_s(\omega)}{\omega-m^2}
+\omega [\SG_v  * \rho_v] +[\SG_s  * \rho_s]
\nn \\
f_2&\equiv&r\frac{m\rho_v(\omega)+\rho_s(\omega)}{\omega-m^2}
+\omega [\SG_s  * \rho_v] +[\SG_v  * \rho_s]
 \quad,
\eea
where $\omega$ is (positive in our metric) time-like momentum $\omega\equiv p^2$  and 
where we have used the abbreviation  for the real functional: 
\be \label{principial}
[\SG * \rho]=P.\int_{m^2}^{\infty} dx\frac{\rho(s)\SG(x)+\SG(s)\rho(x)\frac{s-\mu^2}{x-\mu^2}}{s-x}.
\ee
where P. stands for principal value integration.

Note also, that the fixed renormalization implies the  unique determination of  the propagator  residuum. 
Henceforth, its value ca be obtain obtainable without on-shell differentiation of the self-energy function. Clearly putting $\not p$ t any fixed value the residuum can be easily extracted ( simplifying choice is for instance $\not p=0$ or $\not p=0$).

\section {DSEs in Quantum Electrodynamic}
%*******88888888888888888888888************************88888888888888888888888888888*********************

After the more general introduction we describe  Minkowski formalism which  is necessary in QED  DSEs treatment.
For this purpose we used the conventions already established in the last two previous sections,
the  differences that appear are emphasized. In the end of this section we also review the Euclidean  fermion DSE, its  solution serves us for numerical comparison.  

In general, there is no isolated pole but propagator singularity coincides with the branching point.
The analytical structure of QED fermion propagator was the subject of the initial study 
\cite{KUGO}.
The authors of the paper \cite{KUGO} converted the integral   gap equation 
(for its explicit form see Rel. (\ref{euclid2}) in the  text bellow) to the non-linear differential equation
which has been then solved by  graphical method. In addition, they proceed the backward Wick-rotation
of their equation to the time-like regime of momenta. For a large coupling enough,  they did not find 
 zero in  the inverse electron propagator (for a real $p^2$). This unexpected disappearance of physical branch point
leads the author to the conclusion that confinement should exists even in the Abelian gauge theory.
Having the ultraviolet cut-off $\Lambda$ fixed to the some finite but large  value ($\Lambda>>M(0)$),  the authors of  \cite{KUGO} 
identify the coupling of  the  phase transition to be exactly the one related to the dynamical mass generation, 
i.e. $\alpha=\pi/3 $ in Landau gauge.

In fact, the claim that the Abelian gauge theory has a confinement phase sounds 
strange from the conventional wisdom based on the continuum Abelian gauge theory.
It is clear enough, that the electrons must be a free particles when the coupling is small enough. 
The most recent work can divert this unreliability. It was argued  in the paper
\cite{KONDO} that the Abelian confinement should exist only at the above  certain value
of strong coupling constant, probably where  the photon reveals the dynamical mass.    
Making the quenched approximation, the photon mass generation can not be answered,  but the corresponding
 disappearance of the physical branch point is observed and  confirmed  in our study.
In agreement with our expectation, this transition is observed for the value $\alpha \simeq 1$ 
of QED coupling constant which coincides (up to the numerical accuracy)
 with  the failure of our spectral approach.     

The later analyze  \cite{BLATT} founded  that the electron propagator
(in the  ladder approximation) develops singularity for the all  couplings. 
The results of the paper \cite{MARIS} support partially  
the  conclusion made in  \cite{BLATT} (but, not for all the couplings):
The electron propagator has not only one real branch point, as it is physically expected, but it also 
embodies two other complex conjugated singularities (the position of them entails troubles
with the analytical continuation, see  the discussion in \cite{ROBERTS}). 
To conclude, note that it is generally believed, that 
this unexpected pole's complexity should vanishes for the exact solution 
(particularly when the coupling is small enough). In fact, the existence of the complex branch points  is  questionable and 
the appropriate answer depends on the approximation employed.
Note that, it is not necessary in the contradiction with the result 
of us and with the earlier study of Fukuda-Kugo \cite{KUGO}. It would be only in the case 
if the pure real pole solution does not vanish for the  coupling constants large enough.

In the spectral approach used thorough this  work, the dominant part
of the propagator is driven by the real pole part: $r_f(p^2-m^2)^{-1}$ and  the  changes
coming from the interaction are involved in  continuous part of Lehmann spectrum
\be
\frac{\not\!p\SG_v(\omega)+\SG_s(\omega)}{p^2-\omega+i\epsilon}
\ee
which may but need not to be  analytical at the branch point $p^2=m^2_+$.
Recall at this place the Gauge Technique (zeroth order iteration) result  \cite{CIMENTO} where
in class of covariant gauges the infrared behavior reads
\begin{equation} \label{delb}
S(\not\!p) \simeq \frac{1}{\not\!p-m}\left(\frac{m^2}{p^2-m^2}\right)^
{\frac{\alpha}{2\pi}(\xi-3)}.
\end{equation}
We actually see that the interaction  partially suppressed pole singularity 
for those gauge fixing parameters that are less then  Yennie gauge ($\xi=3$ in Yennie gauge) . 

Furthermore, in  contrast to the YT ,  we are now 
dealing with the gauge theory and the appropriate
renormalization scheme should respect the gauge identity.
Mainly the vacuum polarization tensor should be transverse
\be \label{transv}
q^{\mu}\Pi_{\mu\nu}(q)=0
\ee
and the proper photon-electron-electron vertex should satisfy
Ward-Takahashi identity
\be \label{WTI}
S^{-1}(p)-S^{-1}(l)=(p-l)_{\mu}\Gamma^{\mu}(p,l)\, ,
\ee
which uniquely determined the longitudinal part of the vertex
\cite{BALLCHIU}
\bea  \label{gamaL}
\Gamma^{\mu}_{L}(p,l)&=&\frac{\gamma^{\mu}}{2}\left( A(p^2)+A(l^2)\right)
\nn \\
&+&\frac{1}{2}\frac{(\not p +\not l)(p^{\mu}+l^{\mu})}{p^2-l^2}\left( A(p^2)-A(l^2)\right)
-\frac{p^{\mu}+l^{\mu}}{p^2-l^2}\left( B(p^2)-B(l^2)\right)\, .
\eea
Using the DRs for the function A,B
\bea   \label{selfcon}
A(p^2)&=&1-\int d\alpha 
\frac{\rho_v(\alpha)(p^2-\mu^2)}{(p^2-\alpha+i\ep)(\alpha-\mu^2)}\, ,
\nn \\
B(p^2)&=&1+\int d\alpha 
\frac{\rho_s(\alpha)(p^2-\mu^2)}{(p^2-\alpha+i\ep)(\alpha-\mu^2)}\, ,
\eea
we obtain the integral representation for $\Gamma^{\mu}_{L}$   
\bea  \label{gfirforvertex}
\Gamma_{L}^{\mu}(p,l)&=& 
\gamma^{\mu}\left(1-\int d\omega \frac{\rho_v(\omega)}{\omega-\mu^2}
-\int d\omega \frac{\rho_v(\omega)}{p^2-\omega}
-\int d\omega \frac{\rho_v(\omega)}{l^2-\omega}\right)
\nn \\
&-& \frac{(p^{\mu}+l^{\mu})(\not p+\not l)}{2}
\int d\omega\frac{\rho_v(\omega)}
{(p^2-\omega)(l^2-\omega)}
\nn \\
&-& (p^{\mu}+l^{\mu})
\int d\omega\frac{\rho_s(\omega)}
{(p^2-\omega)(l^2-\omega)}\, .
\eea
From this expression one can read immediately see that the longitudinal part 
of QED vertex is free of any kinematical singularities
and we also see that the only  coefficient of $\gamma^{\mu}$ 
is  explicitly renormalization  point dependent.
 Furthermore, we can note here that this should be 
the whole $\mu$ explicit dependence of the full vertex 
$\Gamma^{\mu}=\Gamma^{\mu}_L+\Gamma^{\mu}_T$, since its transverse part 
(satisfying $(p-l).\Gamma_T=0$) must be finite 
(for more details of (perturbative) MOM renormalization scheme 
used in QED see \cite{SILVERS}).

Substituting the gauge covariant vertex (\ref{gamaL}) into the DSEs for fermion 
and photon propagator we obtain two closed equations that can be solved numerically after the renormalization
(see most recent paper \cite{KIZIL} and references therein)
Adopting rather  standard notation: let the constants $ Z_1$ ,$Z_2$ and $Z_3$ represent the vertex,
fermion wave function and photon wave function respectively then
the renormalization proceeds by the standard way: 
the  Dirac and scalar function $a_0(\mu)$ and $b_0(\mu)$ in unrenormalized  $\Sigma_0$
must be absorbed in $Z_2$ and $Z_m$ renormalization constant by the same 
way as it is happen in the case of Yukawa fermion propagator.
 Furthermore the unrenormalized vacuum polarization $\Pi(\mu)$ should be  absorbed
in the renormalization constant $Z_3$ and the infinity of $\Gamma^{\mu}$ 
is canceled against the constant $1-Z_1$. The  renormalization 
scale used to renormalize vertex and the electron propagator should be 
the same due to the WTI (\ref{WTI}) that requires $Z_1=Z_2$
( Again  we do not explicitly
indicate the dependence on the renormalization scales (i.e., $Z_i=Z_i(\mu)$)).

In our MOM 'imaginary part analysis'  we must self-consistently 
obtain DR (\ref{selfcon}) for fermion function.
Further we should obtain  DR for renormalized photon vacuum polarization

\bea \label{mom}
\pi_R^{\mu\nu}(\mu,q)&=&q^2\left(g^{\mu\nu}-\frac{q^{\mu}q^{\nu}}{q^2}\right)\pi_R^{MOM}(\mu^2;q^2)
\nn \\
\pi_R^{MOM}(\mu^2;q^2)&=&\int\limits_{0}^{\infty}
d\omega \frac{q^2-\mu^2}{(q^2-\omega+i\epsilon)(\omega-\mu^2)}\rho(\omega),
\eea
where its absorptive part reads

\be   
\pi\rho(\omega)=\frac{\alpha_{QED}}{3}(1+2m^2/\omega)\sqrt{1-4m^2/\omega}\Theta(\omega-4m^2)
+ O(e^4)
\ee
with $ \alpha_{QED}= e^2/(4\pi)$, $e$ is a charge of electron, $m$ is  on shell electron mass.
Without making some other approximation the full treatment with the DSEs leads to  the evaluation of 
large number loop integrals  and it also requires a careful numerical treatment due to the presence
of Landau ghost. The complete solution of this problem is relegated to the forthcoming paper \cite{SAULI2}.

Using the bare vertex and quenched approximation  the fermion DSE 
reads

\be          \label{vombat}
Z_2 [\not\!p - m_0]-ie^2\int\frac{d^4k}{(2\pi)^4}G^{\mu\nu}_{0}(k)\gamma^{\mu}
S(p-k)\gamma^{\nu}=0.
\ee
and  straightforward  calculation gives $a(\mu;p^2)=0$. 
This entails equality $A(\mu,p^2)=1$ for all square of momenta \cite{COHEN} if the  condition $A(\mu)=1$ is imposed.
 The  unitary equations are notable  simplified in this case:    
\bea  \label{Tak2}
\SG_v(\omega)&=& \frac{f_1+m(\mu)f_2}{\omega-m^2(\mu)}
\quad ; \quad 
\SG_s(\omega)=\frac{m(\mu)f_1+\omega f_2}{\omega-m^2(\mu)}\, ,
\nn \\
f_1&\equiv&r\frac{m\rho_s(\omega)}{\omega-m^2}
 +[\SG_s  * \rho_s]
\quad; \quad
f_2\equiv r\frac{\rho_s(\omega)}{\omega-m^2}+
[\SG_v  * \rho_s]\, ,
\eea
where the absorptive part of the renormalized self-energy
 $\Im  \Sigma(p)=b(\mu;p^2)=\pi \rho_s(p^2)$ is given by 
\bea  \label{pamfobik}
\rho_s(\omega)=-3\left(\frac{e}{4\pi}\right)^2
\left[r\, m\left(1-\frac{m^2}{\omega}\right)+
\int_{m^2}^{\omega}d\alpha \SG_s(\alpha)\left(1-\frac{\alpha}{\omega}\right)\right].
\eea
Some details of derivation of Rel. (\ref{pamfobik}) are given in the
 part b) of the Appendix A. 

In order to make a careful and constructive comparison between Minkowski approach 
presented in this work and the standard Euclidean formulation  we 
should compare with some known Euclidean results presented in the literature\cite{KUGO},\cite{HAWE}.
In space-like region the Eq. (\ref{vombat}) is transfered to 
\bea
B(x)&=&m_0+\frac{3\alpha}{4\pi}\int\limits_0^{\infty} dy K(x,y) \frac{B(y)}{y+B^2(y)}
\nn \\
K(x,y)&=&\frac{2y}{x+y+\sqrt{(x-y)^2}}
\eea
where Wick rotation 
 and angle integration have been done and  where we have used  $Z_2=1$,
$x\equiv p^2_E=-p^2$,$y\equiv k^2_E=-k^2$. The renormalized equation then reads
\bea     \label{euclid}
B(\zeta,x)&=&m(\zeta)+\frac{3\alpha}{4\pi}\int\limits_0^{\infty} dy V(\zeta,x,y) \frac{B(\zeta,y)}{y+B^2(\zeta,y)}
\nn \\
V(\zeta,x,y)&=&K(x,y)-K(\zeta,y)
\eea  
where $\zeta$ is  square of renormalization scale $-\mu^2$. Choosing the scale $\zeta$ to be zero and further 
scaling the renormalized $m(\zeta) $ mass as $m(0)=1$ leads to the particularly simple expression for (\ref{euclid})
\bea  \label{euclid2}
B(0,x)&=&1+\frac{3\alpha}{4\pi}\int\limits_0^{x} dy \left(\frac{y}{x}-1\right) \frac{B(0,y)}{y+B^2(0,y)}
\eea
(for 'massive photon' case see Eq. (3.14) in the work \cite{KUGO}).
Stressed here, that due to the Landau gauge the solution of Eq.(\ref{euclid}) is only 
 slightly deviating from the solution with Curtis-Pennington vertex implemented \cite{HAWE}.
From the paper \cite{WILDA}  the asymptotic behavior of dynamical mass is known  analytically.
Henceforth, any reliable solution of electron DSE must behaves at ultraviolet like
\be
M(p_E^2)=M(\zeta)\left(\frac{p^2_E}{\zeta}\right)^s.
\ee

\section{Numerical Solution and Results}

\subsection{Yukawa theory}

The resulting coupled nonlinear integral equations (\ref{Takk}),(\ref{pyjon}) for the functions $\SG_{v,s}$
and $\sigma_{\phi}$ require the knowledge about the
value of  physical masses (\ref{defonshell}), (\ref{klokan}), the propagators residua (\ref{rezion}),(\ref{rezid})
and the complete knowledge of the absorptive parts of self-energies (\ref{substrakce}),(\ref{zadny}),(\ref{becko}).
The equations  have been solved  by the method of numerical iteration which  seems to be particularly useful for this purpose.
The one loop perturbation theory result is used as the  zeroth order of this iteration.
Then, several hundred of iteration steps have been proceed to achieve a stable  solution.
For purpose of numerical integration we  choose the Gaussian quadrature method. 
Taking a reasonable number of the integration mash points 
and adopting the principal value integration described in the Appendix B,
then the whole numerical  procedure is  rather stable against the
changing of mesh points density and their number as well as.
     
The all presented results for YT are  evaluated for the zero value of
renormalization scale $\mu^2=0$ which choice is common for  propagators corresponding 
with both the particle of YT. 
The physical mass is  usually the best known experimentally and we face that this is our case.
Let us assume that  experimentalists (living in our toy model world)
found their values: $m_{\phi}=0.15m$, $m=1$.
In this case, we are enforced to calculate $m(\mu)$ and $m_{\phi}(\mu)$ from the Rels.
(\ref{defonshell}), (\ref{klokan}) which procedure does not cause any troubles in our iteration method. 
The above described procedure has  clear numerical advantage:  
the branching points lie at the values of momenta which are fixed at each iteration step,
but it has also disadvantage: the comparison with some results obtained in the  Euclidean
formalism is not so straightforward (note, $m(\mu)$ is fixed for some space-like $\mu^2$).
To make the most accurate comparison with Euclidean result, 
we should fix  the renormalization scale and the renormalized masses at the same
values that are used in the Euclidean DSEs. We prefer the first scheme in YT since
the appropriate Euclidean solution is not published elsewhere
(but we use the second approach in  QED case).    
  
Fig.1 shows  the absorptive parts of self-energies, i.e., the functions 
$\rho_v,\rho_s$ for fermion and  $\rho_{\phi}$ for pseudoscalar.
They are plotted 
for the following values: $\lambda=0.1,0.2,0.3$ of the coupling strength  defined 
as $\lambda=\frac{g^2}{4\pi}$. For a large momenta the  spectral function $\rho_{\phi}(x)$
grows linearly with the square of momenta which  is a consequence of 
 quadratic divergence in $\Pi_0$. Due to this, what we actually plot is the rescaled 
function $\rho_{\phi}(x)/x$. All the functions start to be non zero from their 
perturbative thresholds-  $\Im  A(s),\Im  B(s)$ from $s=(m_{\phi}+m)^2$ and $\Im  \Pi(s)$ 
from $s=4m^2$. Because the negative parity of the field $\Phi$  the functions 
$\rho_s,\rho_{\phi}$ (and $\SG_s,\SG_{\phi}$) are positive while 
$\rho_v$ ($ \SG_v$) is negative.
The quadratic divergence of the unrenormalized 
self-energy $\Pi$  leads to the quadratic dependence (+corrections) of the renormalized $\Pi$.
(in fact the recent  models (see for instance \cite{COHEN2},\cite{LOGAN})
of particle interactions attempted to  avoid of  quadratic divergences that
necessarily follows from Yukawa sector of Standard model)
The quantity $Q^{1/2}$ of dimension [mass$]^1$ defined with the help of  eq.
$ Q(p^2)=m_{\phi}^2(\mu)+\Pi(\mu;p^2)$ can not be called  the 
dynamical boson mass, since $Q$ becomes negative above the certain value of Euclidean momenta.
The same happens at the time-like axis for $\Re Q(p^2)$ (while $\Im  Q(p^2)=\pi\rho_{\phi}$ is positive 
as it is clear from  Fig.1).      
The appropriate behavior of the function $Q$ is displayed in  Fig.2. 
The  momentum dependence of the fermion functions $A,B$ is dominated
by the (perturbative) logarithm of $p^2$. Indeed, the Yukawa fermion propagator is almost given by its free form
corrected by small perturbation.
 Fig. 3 and  Fig.4 display the momentum dependence  of the function $A,B$ for 
space-like momenta and time-like momenta respectively.
Although, the smallness of the ratio $m_{phi}/m$ is motivated by the desired enhancement of 
fermion self-energy and subsequent suppression of pseudoscalar one, nevertheless, we still see
that the fermion-meson loop becomes irrelevant perturbative contribution for 
all studied couplings. We display the  dynamical mass $M=B/A$ in the infrared domain  in the Fig.5. 
 since this is the mainly  interesting regime of momenta.
The appearance 
of the critical coupling $\lambda_c$ is the consequence of the 
quadratic momentum dependence of pseudoscalar self-energy.
Its value slightly depends  on the  numerical cut-off $\Lambda$ used
in our unitary equation, noting that within our numeric ($\Lambda^2=10^7m^2$) the solution 
of unitary equations fails at  $\lambda_c\simeq 0.3$.

\subsection{ QED fermion propagator}

In contrast to the model  discussed previously, the strong coupling QED 
is less driven by the perturbation theory (Recall here the famous paper
\cite{COLEMAN} dealing with scalar electrodynamics).Before presentation of our numerical solution, let us 
recollect the main results:

1. Comparing the  electron propagator obtained
from the solution of  the Unitary Equations (UEs) (\ref{Tak2})  with the propagator calculated in the Euclidean formalism we find
that they never exactly agree. The exception is the case of very small coupling $\alpha\simeq 10^{-2}$,
where both approaches seem to be equivalent  (up to the numerical errors). Stressed here, that in this case,
they are almost indistinguishable form the  perturbation theory result.

2. Previous statement is valid  also in the case when the small  photon mass $\lambda$ is introduced.
The obtained results are then  slightly changed quantitatively and the disagreement discussed in 1. is somehow small
(the Euclidean$\times$Minkowski results numerically agree, only when $\lambda\simeq m$,
 which is the case, we are not interested in).
It leads us to the  conclusion, that this discrepancy do not fully follows from the masslessness of the photon 
and from the appropriate subsequent coincidence of the electron propagator singularity  with the branch point.
Note, that the last property is not simply fulfilled for nonzero $\lambda$.              

3. The Minkowski (spectral) solution exists only for subcritical regime of $\alpha_{QED}$, whilst, as it is well
known  the Euclidean solution  can be easily  find for both the sub- and super-critical couplings.
 To be more precise, we have not found any solution of UEs
when the coupling was larger then $\alpha_{c.s.}=0.95$ (c.s.=critical spectral).
Further, making an estimate of  the momentum DSE solution   (not assuming spectral decomposition) we have found strong evidence for confinement
for the coupling larger then $\alpha=0.91\pm0.1$. 
 (Remind,that the critical coupling in QED is defined such that
$B(p^2)=0$ for $m_0=0$, (chiral symmetric phase) for $\alpha<\alpha_c$ and the solution for $B(p^2)$ is non-zero and finite
for $\alpha>\alpha_c; m_0=0$ when momentum cut-off $\Lambda$ is implemented 
(without finite $\Lambda$  the function $B$ tends to diverge). Its value is known from the quenched, rainbow study, where $\alpha_c=\frac{\pi}{3}$
, more sophisticated studies do not deviate significantly from this value ).      

The UEs (\ref{Tak2}) complemented by the equations for the absorptive part $\rho_s$ and for the residuum have been
solved by the method of iterations. In order to have an infrared behavior under the control and in order to see the
effect of the vector boson mass  we also implement the small mass parameter $\lambda$ into  the photon propagator:
\be
G_o^{\mu\nu}(k)=\frac{-g^{\mu\nu}+k^{\mu}k^{\nu}/k^2}{k^2-\lambda^2+i\ep}
\ee.
Restricting to the small photon mass case, the Dirac part  coefficient function $A$ is approximated as $ A(p^2)=1 $, 
i.e. as it would be in the massless case. 
The expression for  function $\rho_s$ is then slightly change (see Appendix A), noting here that the limit
$\lambda\rightarrow 0$ is smoothly achieved as 
it follows from the appropriate relations (\ref{hmotny}) presented in the Appendix A.
The solutions with several different photon masses have been obtained. 
Then, the dynamical mass $M=B$ at space-like momenta is calculated from the  absorptive 
functions $\rho_s,\sigma_s,\sigma_v $, which are  primary solutions here.      
We can see from the Fig.6. that the effect of the small  photon mass is almost  irrelevant
even if the coupling is relatively large ($\alpha=0.6$).
The space-like renormalization scale is chosen to be large 
$\mu^2=-10^8$ when compared with renormalized mass  $m^2(\mu)=400^2$, which choice corresponds with the   Euclidean solution 
already presented in the work \cite{WILL2}. The appropriate Euclidean solution of Eq. (\ref{euclid}) 
is added for the comparison.  The same (in)dependence on the parameter $\lambda$ is observed
for any choice of $\mu,m(\mu)$.    
In order to achieve good numerical stability of UEs some very small photon mass is always used in.
The presented results in this work are calculated with $\lambda=10^{-3}m$.

Solutions were obtained for the Fukuda-Kugo equation (FKE) (\ref{euclid2}) for the couplings from $0.01$ to $2.0$.
The expected damping \cite{KUGO,HAWE}  of dynamical mass to its negative values was observed for supercritical couplings $\alpha>\pi/3$.
Using  the same renormalization choice $M(0)=1$ 
the  UEs were solved. The resulting solution for $M$ is plotted for the following couplings: $0.2,0.4,0.6$ and compared with the 
FKE solution.  All these solutions are plotted against the space-like momentum in Fig.7.
From this we can see the relatively large discrepancy between
the results of FKE and UEs.
 Making great  numerical effort we have found the main cause numerically. Let us assume that 
 the  residuum and the principal value integral integrations that appear in UE  are somehow over-estimated due to the infrared
enhancement of the Lehmann weights. Let us introduce small coupling constant dependent infrared cutoff $c*m^2$ to the UEs:
\bea  \label{modified}
\SG_v(\omega)&=& \frac{f_1+m(\mu)f_2}{\omega-m^2(\mu)}
\quad ; \quad 
\SG_s(\omega)=\frac{m(\mu)f_1+\omega f_2}{\omega-m^2(\mu)}
\nn \\
f_1&\equiv&r\frac{m\rho_s(\omega)}{\omega-m^2}
 +[\SG_s  * \rho_s]\Theta(\omega-m^2(1+c))
\nn \\
f_2&\equiv& r\frac{\rho_s(\omega)}{\omega-m^2}+
[\SG_v  * \rho_s]\Theta(\omega-m^2(1+c))
\nn \\
r&=&\left(1+m\frac{m-m(\mu)}{m(\mu)m-\mu^2}\right)
\left(1+ \int d\omega \frac{\mu^2 \SG_v(\omega)-m(\mu)\SG_s(\omega)}{\omega-\mu^2}\right)\Theta(\omega-m^2(1+c))
\nn \\
c&=&\frac{\alpha}{2\pi} \, .
\eea
Note that taking $c=0$ then the  modified unitary equations (MUEs)
 (\ref{modified}) clearly reproduce the originally UEs.
Looking at the solutions of MUEs we see that their solutions are
 in the reasonable agreement  with the solutions 
of FKE. 
Fig.8 displays the mass behavior at the time-like regime  
obtained for the same choice of renormalization mass as previous.
The maximum corresponds correspond with the  mass-shell point . 
Note that their  existence is in contradiction with the assumption of confinement \cite{KUGO}. 
Although, we are pretty sure that the solution must look like as in this figure, in order to assure, we solve
the time-like continued equation of Fukuda-Kugo. We  were not able to find the exact and fully complex
 solution numerically, however,  we made a sophisticated estimate of a real part of $B(p^2)$ by the  principal value integration. 
It is achieved  by the standard numerical introduction of a finite epsilon,
 here it  is chosen to be a fraction of pole mass and  presented solutions are obtained with the choice
$\epsilon=0.03m^4$,  i.e., the singular integrand is approximated as
\be  \label{epsik}
P.\frac{1}{z-B^2(z)}\rightarrow \frac{z-B^2(z)}{(z-B^2(z))^2+\epsilon}
\ee
where $z$ is a time-like momentum $z=p^2$.

The soft coupling results are added to the Fig.8 and fully agree with our expectation: 
the branch point always exists because  the mass function $B^2(x)$ always
cuts the graph  of the function $y(x)=x$. This statement is valid for QED with subcritical couplings $\alpha<0.915$.   
At most, we expect about ten percentage deviation from the exact value of  $\alpha_{crit}$ which should be a consequence
of the numerical method weakness. 
Furthermore, we can see from the Fig.8 that FKE solution for  $\Re B$ reasonably agree with the solutions of MUEs,
especially when the coupling is small enough. But we have rather different situation in strong coupling QED. The maximum 
of the mass function disappear and the mass function never cuts  the graph $\sqrt{x}$. {\it The supercritical coupling quenched QED 
is a confining theory, the electron propagator has not a  branch point and is free of any  singularities}.
In this case,  
$\epsilon$ introduced above can be safely limited to the zero value and we actually omitted it when FKE was solved for 
$\alpha>\alpha_{crit}$. 
 The results are plotted
in Fig.9 where we also add some subcritical solutions for better comparison.
The subcritical solutions have been obtained with the help of time-like continued FKE and 
there is no spectral solutions presented in this figure (the value of infrared cut-off $c$ makes the MUEs untrusted).       
           
The program using the Euclidean equation (\ref{euclid}) was run  with several coupling $\alpha$ from
$0.2$ to $8.0$ (!), and with the  renormalization choice $m(-100)=10$.
 Whenever $\alpha$ coincides
with the one published in \cite{HAWE}, then the results here and the ones obtained in \cite{HAWE} 
should be  numerically identical. Again, we see  for supercritical couplings, there is a small region where the dynamical mass is negative.
When the coupling constant is extremely large (say $\alpha=8$) the negative damping becomes a
relatively fast oscillation around  the zero mass axis. This feature should be in  agreement with 
\cite{HAWE} but in a subtle disagreement with the paper \cite{KUGO}, where the  oscillation appears to be purely positive.
 The solution of $B$ in space-like regime is presented in  Fig.10.
The solutions of MUEs and of the Euclidean momentum DSE are compared only in the space-like regime. The MUEs solutions are presented in 
 in Fig. 11 for time-like momenta.    

An unexpected feature of the spectral equations is that for any set $\{\alpha>0.9,\mu^2,m(\mu)\}$
the absorptive self-energy becomes oscillating around the zero axis. 
This behavior was verified to be insensitive to number of grid points, but was not stable against the infrared cut-off introduced above.
In fact, the positions of minima's and maxima's are slowly walking when the cut-off $c$ is varied by hand. The significance of this for
spectral representation and its connection to QED is not completely understood. One possibility is that it may be signal the failure
of Lehmann representation. The examples of this feature is shown in Fig.12. 
It should be emphasized here, that all the results for $M(p^2)$ obtained by solving of  MUEs
and presented in this work have been calculated from non-oscillating, smoothed (soft coupling constant) $\rho_s$'s.

\section{Conclusions and outlook}

In this article we convert the Dyson-Schwinger equations for propagators
to the real equations for their Lehmann weights.
The possibility to do this, is  based on the assumption of the existence of  Green's function integral representation which should be valid in nonperturbative regime too.
They renormalized equations have  been solved without some unwanted linearization or angle approximation.
 This   is the novelty and significant
advantage in comparison with  the usual manner commonly used in the literature.
The developed approach has been tested on  two models.
The  QED electron propagator has been compared with the results obtained in conventional Euclidean formalism.
We found  certain discrepancy between these approaches which appears to be rather large
when the coupling approaches its critical value.
Furthermore, the solution obtained within the help of  Lehmann representation
is fully absent for supercritical couplings regime of QED.
In this case the physical pole propagator singularity disappear and we argue that ladder
QED does not describe free electrons(positrons) at all.
 
\begin{center}{\bf Acknowledgments}\end{center}

Author is very grateful to I.Kavkov\'a , R.Derm\'{\i}\v{s}ek and A.Ciepl\'y for the 
careful reading of the manuscript. I also thank to prof. R.Delbourgo 
who point my attention on some existing literature.
This research was supported by GA \v{C}R under Contract n.202/00/1669.

%%%%%%%%%%%%%%%%%%%%%%%%%%%%%%%%%%%%%%%%%%%%%%%%%%%%%%%%%%%%%%%%%%%%%%%%%%

\appendix

\section{Evaluation of the off-shell dispersion relations}

In this Appendix we derive the appropriate DR's for the  proper Green's functions of  YT.  
As a first step  we  analyze the fermion self-energy.

\subsection{Yukawa fermion self-energy}

 As follows from (\ref{zuzu}) the general
expression which has to be considered has the following structure:
\bea          \label{sigi}
\Sigma_0(\not\!p)&=&-ig^2\int\frac{d^4l}{(2\pi)^4}\frac{\gamma^{5}
(C_a \not\!l+C_b) \gamma^{5}}{(l^2-\alpha+i\epsilon)((l-p)^2-\beta+i\epsilon)}
\nn \\
&=&-ig^2\int\limits_{0}^{1}dx\int\frac{d^4l}{(2\pi)^4}
\frac{-C_a\not\!p(1-x)+C_b}{[l^2+p^{2}x(1-x)-\alpha x-\beta(1-x)+i\epsilon]^2} \quad ,
\eea
where the whole prefactors are absorbed  into  the formal symbols $C_{a,b}$ .
 Written in the terms of continuous   functions $\SG_{v,s}$ and Lehmann weight $\SG$ for pseudoscalar,
the symbol $C_a$ and $C_b$ are identified as
\bea
C_a&=&\int d\alpha\left\{ \delta(\alpha-m^2)+
\SG_v(\alpha)\right\}   
\int d\beta \left\{\delta(\beta-m_{\phi}^2)+\SG(\beta)\right\}
\nn \\
C_b&=&\int d\alpha\left\{ \alpha^{1/2}\delta(\alpha-m^2)+
\SG_s(\alpha)\right\}   
\int d\beta \left\{\delta(\beta-m_{\phi}^2)+\SG(\beta)\right\}
\eea
Noting that they should be collocated in front of the fractions in (\ref{sigi}).

Observing the Dirac structure, the unrenormalized functions $a_0,b_0$ 
 can be simply identified. 
After making subtractions (\ref{abdisperze}) we can arrive at their renormalized forms 
\bea           
a(\mu;p^2)&=&
ig^2C_a\int\limits_{0}^{1}dxdy\int\frac{d^4l}{(2\pi)^4}
\frac{(1-x)^2x2(p^2-\mu^2)}{\left[l^2+(p^2-\mu^2)x(1-x)y+\mu^2x(1-x)-O+i\epsilon\right]^3}
\nn \\
\\
b(\mu;p^2)&=&
-ig^2C_b\int\limits_{0}^{1}dxdy\int\frac{d^4l}{(2\pi)^4}
\frac{2(1-x)x(p^2-\mu^2)}{\left[l^2+(p^2-\mu^2)x(1-x)y+\mu^2x(1-x)-O+i\epsilon\right]^3}
\nn
\eea
where $O=\alpha x-\beta(1-x) $.
Integrating over the loop momentum and making substitution $y\rightarrow \omega$
\be  \label{zvjozda}
\omega=\mu^2-\frac{\mu^2}{y}+\frac{O}{x(1-x)y} ,
\ee
  we obtain  the DR's 
\bea \label{osp}
a(\mu;p^2)&=&-C_a\left(\frac{g}{4\pi}\right)^2
\int\limits_{0}^{1}dx\int\limits_{\frac{O}{x(1-x)}}^{\infty}d\omega
\frac{1-x}{\omega-\mu^2}\frac{p^2-\mu^2}
{(p^{2}-\omega+i\epsilon)}
\nn \\
b(\mu;p^2)&=&C_b\left(\frac{g}{4\pi}\right)^2
\int\limits_{0}^{1}dx\int\limits_{\frac{O}{x(1-x)}}^{\infty}d\omega
\frac{p^2-\mu^2}{\omega-\mu^2}\frac{1 }
{(p^{2}-\omega+i\epsilon)}  \quad.
\eea
Using the following definition of X functions
\bea \label{easily}
X(\omega;\alpha,\beta)&=&\int\limits_{0}^{1}dx\Theta\left(\omega-\frac{O}{x(1-x)}\right)
\nn \\
X_1(\omega;\alpha,\beta)&=&
\int\limits_{0}^{1}dxx\Theta\left(\omega-\frac{O}{x(1-x)}\right)
\eea
we can rewrite DR(\ref{osp}) into the more familiar form:
\bea \label{misjapec}
a(\mu;p^2)&=&-C_a\left(\frac{g}{4\pi}\right)^2\int d\omega
\frac{X_1(\omega;\beta,\alpha) (p^2-\mu^2)}
{(\omega-\mu^2)(p^{2}-\omega+i\epsilon)}
\nn \\
b(\mu;p^2)&=&C_b\left(\frac{g}{4\pi}\right)^2\int d\omega
\frac{X(\omega;\alpha,\beta) (p^2-\mu^2)}
{(\omega-\mu^2)(p^{2}-\omega+i\epsilon)} \quad ,
\eea
where we made use of the relation
\be \label{symmetry}
 X(\omega;\alpha,\beta)- X_1(\omega;\alpha,\beta)=
X_1(\omega;\beta,\alpha) \quad .
\ee

Explicit integrations in (\ref{easily}) give the following formulas for $X$ functions
\bea \label{becka}
X(\omega;\alpha,\beta)&=&\frac{\lambda^{1/2}(\alpha,\omega,\beta)}{\omega}
\Theta\left(\omega-(\alpha^{\frac{1}{2}}+\beta^{\frac{1}{2}})^2\right)
\nn \\
X_1(\omega;\alpha,\beta)&=&\frac{\lambda^{1/2}(\alpha,\omega,\beta)}{2\omega}
\left[1+\frac{\beta-\alpha}{\omega}\right]
\Theta\left(\omega-(\alpha^{\frac{1}{2}}+\beta^{\frac{1}{2}})^2\right)\, ,
\eea
where $\lambda$ is  the triangle Khall\'en  function 
$\lambda=(\omega-\alpha+\beta)^2-4\omega\beta$.

After the explicit introduction of the prefactors $C_{a,b}$ we can write down
the absorptive part of self-energy functions $a,b$:
\bea \label{roa}
\rho_a(\omega)&=& \frac{-g^2}{(4\pi)^2}
\left[r_f r_{\phi} X_1(\omega;{m_{\phi}}^2,m^2)
+mr_{\phi}\int\limits_{(m+m_{\phi})^2}^{\infty}d\alpha
\SG_v(\alpha)X_1(\omega;m_{\phi}^2,\alpha)\right.
\nn \\
&+&r_f\int\limits_{4m^2}^{\infty} d\beta
\rho_{\phi}(\beta)X_1(\omega;\beta,m^2)
\nn \\
&+&\left.\int\limits_{4m^2}^{\infty} d\beta\int\limits_{(m+m_{\phi})^2}^{\infty}d\alpha
\rho_{\phi}(\beta)\SG_v(\alpha)X_1(\omega;\beta,\alpha)\right],
\eea
\bea \label{rob}
\rho_b(\omega)&=& \frac{g^2}{(4\pi)^2}
\left[r_{\phi}r_f mX(\omega;m_{\phi}^2,m^2)
+r_{\phi}\int\limits_{(m+m_{\phi})^2}^{\infty}d\alpha
\SG_s(\alpha)X(\omega;m_{\phi}^2,\alpha)\right.
\nn \\
&+& r_f\int\limits_{4m^2}^{\infty} d\beta
\rho_{\phi}(\beta)X(\omega;\beta,m^2)
\nn \\
&+&\left.\int\limits_{4m^2}^{\infty} d\beta\int\limits_{(m+m_{\phi})^2}^{\infty}d\alpha
\rho_{\phi}(\beta)\SG_s(\alpha)
X(\omega;\alpha,\beta)\right].
\eea
As follows from the properties of function $X$ the  various terms in (\ref{roa}),(\ref{rob})
start to be nonzero  from  different values of $\omega$. This knowledge is particularly
useful when calculated numerically. Indeed after the inspection of  $X(\omega;\alpha,\beta)$ 
we can find the threshold at $\omega=(\sqrt{\alpha}+\sqrt{\beta})$.
For instance the subtresholds values $(m_{\phi}+m)^2; (2m_{\phi}+m)^2; 9m^2; (m_{\phi}+3m)^2$ 
correspond to the terms in Eq. (\ref{rob}) at given order.

\subsection{ Yukawa pseudoscalar self-energy}

Let us find the DR for  self-energy $\Pi$.
The loop integral  has the  form: 
\bea   \label{prcina}
\Pi(p^2)&=&i\int\frac{d^4l}{(2\pi)^4}Tr\frac{\gamma_5
[\not\!l\bar{\SG}_v(\alpha) +\bar{\SG}_s](\alpha)\gamma_5
[(\not\!l-\not\!p)\bar{\SG}_v(\beta) +\bar{\SG}_s(\beta)]}
{(l^2-\alpha+i\epsilon)((l-p)^2-\beta+i\epsilon)}
\nn \\
&=&4i\int\frac{d^4l}{(2\pi)^4}
\left[\frac{-\bar{\SG}_v(\alpha) \bar{\SG}_v(\beta)  
}{((l-p)^2-\beta+i\epsilon)}\right.
\nn \\
&+&\left.\int\limits_0^1 dx \frac{(-\alpha+p^2(1-x)) \bar{\SG}_v(\alpha) \bar{\SG}_v(\beta)  
+ \bar{\SG}_s(\alpha) \bar{\SG}_s(\beta)  }{l^2+p^2x(1-x)-O}\right] ,
\eea
where the shift $l\rightarrow l+p(1-x)$
is made at the second integral and $O=\alpha x-\beta(1-x) $.
For purpose of brevity we omit the spectral integration over the variables 
$\alpha$ and $\beta$. 

The  renormalization  (\ref{subtrakce}) proceeds by direct 
subtracting of the first two term in Taylor expansion of unrenormalized 
self-energy (\ref{prcina}). To make this explicitly  we introduce 
shorthand notation  $U(w,l,O)=(l^2-w^2(1-x)x-O+i\epsilon) $ .
After a little algebra the  renormalized quantity corresponding to  
(\ref{prcina}) can be   evaluated as
\bea  \label{veverka}
\Pi(\mu;p^2)&=&4i\int\frac{d^4l}{(2\pi)^4}\int\limits_0^1 dxdy 
\biggl[( -\alpha\bar{\SG}_v(\alpha)\bar{\SG}_v(\beta)
\nn \\
&+& \bar{\SG}_s(\alpha)\bar{\SG}_s(\beta))\left(\frac{1}{U^2(p,l,O)}
- \frac{1}{U^2(\mu,l,O)}
+\frac{2x(1-x)(p^2-\mu^2)}{U^3(\mu,l,O)}\right)
\nn \\
&+&\bar{\SG}_v(\alpha)\bar{\SG}_v(\beta)
 \left(\frac{p^2(1-x)}{U^2(p,l,O)}-
\frac{\mu^2(1-x)}{U^2(\mu,l,O)}+ \frac{2x(1-x)^2\mu^2(p^2-\mu^2)}
{U^3(\mu,l,O)}\right)\biggr].
\eea
Matching  terms in the first line of (\ref{veverka}) together 
we can rewrite them into the familiar DR
\be
\frac{4g^2}{(4\pi)^2}\int d\omega \frac{X(\omega;\alpha,\beta)
(-\alpha\bar{\SG}_v(\alpha)\bar{\SG}_v(\beta)+ \bar{\SG}_s
(\alpha)\bar{\SG}_s(\beta))
(p^2-\mu^2)^2}
{(\omega-\mu^2)^2(p^2-\omega+i\epsilon)} \quad ,
\ee
where function $X$ was already introduced in (\ref{becka})

The  terms in the second brackets of the second line of (\ref{veverka})  
can be matched together in the following fashion:
\bea \label{blecha}
&&\int\limits_0^1dy\left[-\frac{2(p^2-\mu^2)^2x(1-x)^2}
{( U(p,l,O)y+ U(\mu,l,O)(1-y))^3}\right.+
\nn \\
&&\left.\left(2x(1-x)^2\mu^2(p^2-\mu^2)\right)
\left(\frac{1}{U^3(\mu,l,O)}-
\frac{1}{( U(p,l,O)y+ U(\mu,l,O)(1-y))^3}\right)\right]=
\nn \\
&&\int\limits_0^1dy\left[-\frac{(p^2-\mu^2)^2(1-x)}
{y\left(p^2-\mu^2+\frac{\mu^2}{y}-\frac{O}{x(1-x)y}\right)}\right.
\nn \\
&&+\left.\frac{(p^2-\mu^2)^2(1-x)\mu^2}
{\left(\mu^2-\frac{O}{x(1-x)y}\right) 
\left(p^2-\mu^2+\frac{\mu^2}{y}-\frac{O}{x(1-x)y}\right)}\right].
\eea
Adding omitted prefactors, integrating over loop momentum and making 
the substitution  (\ref{zvjozda}) in (\ref{blecha})  gives rise to the DR:
\be
\frac{4g_a^2}{(4\pi)^2}\int d\omega \frac{(X(\omega;\alpha,\beta)-
X_1(\omega;\alpha,\beta))\omega(p^2-\mu^2)^2}
{(\omega-\mu^2)^2(p^2-\omega+i\epsilon)}.
\ee
Putting  all together and using the relation between $X$ and $X_1$
we can finally write down  the  DR for self-energy $\Pi$:
\bea
\Pi(\mu;p^2)&=&\int d\omega 
\frac{\rho_{\phi}(\omega)(p^2-\mu^2)^2}{(\omega^2-\mu^2)^2
(p^2-\omega+i\epsilon)}
\nn \\
\rho_{\phi}(\omega)&=&\frac{4g_a^2}{(4\pi)^2}
\int d\alpha \int d\beta \left[\left(-\alpha \bar{\SG}_v(\alpha)\bar{\SG}_v
(\beta) + \bar{\SG}_s(\alpha)\bar{\SG}_s(\beta)\right)X(\omega;\alpha,\beta)\right.
\nn \\
&+&\left.\omega X_1(\omega;\beta,\alpha) \bar{\SG}_v(\alpha)\bar{\SG}_v(\beta) \right].
\eea

For the purpose of numerical solution it is necessary to extract
the singular parts of $\bar{SG}'s$ explicitly. 
After some trivial  manipulation the appropriate formula for $\rho_{\phi}$  reads
\bea \label{vysledek}
&&\rho_{\phi}(\omega)=\left(\frac{g}{2\pi}\right)^2
\left[r_f^2\frac{\omega}{2}\sqrt{1-\frac{4m^2}{\omega}}\Theta(\omega-4m^2)\right.
\nn \\
&+&2r_fm\int\limits_{(m+m_{\phi})^2}^{\infty}d\beta
\left(\frac{1}{2}(\omega-m^2-\beta)\SG_v(\beta)+m\SG_s(\beta)\right)
X(\omega;m^2,\beta)
\nn \\
&+&\int\limits_{(m+m_{\phi})^2}^{\infty}d\alpha \int\limits_{(m+m_{\phi})^2}^{\infty}d\beta 
\left(\frac{1}{2}(\omega-\alpha-\beta)\SG_v(\alpha)\SG_v(\beta)\right.
\nn \\
&+&\left.\left.\SG_s(\alpha)\SG_s(\beta)\right)
X(\omega;\alpha,\beta)\right],
\eea

noting that the first line of (\ref{vysledek}) corresponds with the one loop perturbative contribution.

\section{ Quenched, rainbow, Landau gauge QED }

It is shown, that with a massless photon the fermion self-energy function $b$
satisfies DR (\ref{abdisperze}) while the function $a$ 
is exactly zero. In the end of this section we review the
results for the case of massive photon. These are used in the numerical study described in the main text.
 
First we will deal with massless photon case ($\lambda=0$), where our approximate electron self-energy
\be          \label{sigma0}
\Sigma_0(\not\!p)=ie^2\int\frac{d^4k}{(2\pi)^4}G^{\mu\nu}_{0}(k)\gamma^{\mu}
\int d{\alpha}\frac{\bar{\SG}_v(\alpha)(\not\!p-\not\!k)+\bar{\SG}_s(\alpha)}{(p-k)^2-\alpha+i\epsilon}\gamma^{\nu}\, ,
\ee
requires one subtraction due to the presence of logarithmic ultraviolet divergence.
As it is sometimes usual, we fist regularize (\ref{sigma0}), then the subtraction will follow.
For this purpose  we use the old-fashion Pauli-Villars 
regularization technique which leads to the following  regularized result: 
\bea
\Sigma_{\Lambda}(\not\!p)&=&-\frac{e^2}{(4\pi)^2}
\int d{\alpha}\int\limits_{0}^{1}dx
\left[(\bar{\SG}_v(\alpha)\not\!p(4x-2)+3\bar{\SG}_s(\alpha))\right.\, ,
\nn \\
&\times&\ln{\left(\frac{-p^{2}x(1-x)+\alpha x+
\Lambda^2(1-x)}
{-p^{2}x(1-x)+\alpha x}+i\epsilon\right)}
\nn \\
&+&\left.\frac{2\bar{\SG}_v(\alpha)\not\!pp^2x^2(1-x)}{p^{2}x(1-x)+\alpha x}\right]\, ,
\eea
where $\Lambda$ represents Pauli-Villars regulator.
Making the aforementioned  subtraction
gives the renormalized functions $a,b$     

 \bea     \label{B0}      
a(\mu;p^2)&=&- \int d{\alpha} \frac{\bar{\SG}_v(\alpha)e^2}{(4\pi)^2}\left[\int\limits_{0}^{1}dx
(4x-2)\ln{\left(\frac{-\mu^2(1-x)+\alpha}
{-p^{2}(1-x)+\alpha}+i\epsilon\right)}\right.
\nn \\
&+&\left.\frac{2p^2}{-p^{2}(1-x)+\alpha}
-\frac{2\mu^2}{-\mu^{2}(1-x)+\alpha}\right]
\nn \\
b(\mu;p^2)&=&-\int d\alpha \frac{3\bar{\SG}_s(\alpha)e^2}{(4\pi)^2}\int\limits_{0}^{1}dx
\ln{\left(\frac{-\mu^2(1-x)+\alpha}
{-p^{2}(1-x)+\alpha}+i\epsilon\right)}.
\eea
where $\mu^2$ is   space-like renormalization scale.
The per-partes integration  of logarithm terms gives
\bea \label{sig0}
a(\mu;p^2)&=&0
\nn \\
b(\mu;p^2)&=&\int d\alpha \frac{3\bar{\SG}_s(\alpha)e^2}{(4\pi)^2}\int\limits_{0}^{1}dx
\frac{x}{(1-x)}\frac{p^2}
{(p^{2}-\frac{\alpha}{1-x}+i\epsilon)}-(p^2\rightarrow\mu^2)
\nn \\
&=&-\left(\frac{3e}{4\pi}\right)^2\int d\alpha\int\limits_{\alpha}^{\infty}d\omega
\frac{3\bar{\SG}_s(\alpha)(1-\frac{\alpha}{\omega})(p^2-\mu^2)}
{(\omega-\mu^2)(p^{2}-\omega+i\epsilon)}\, ,
\eea
Thus absorptive part of self-energy $\pi\rho$ is given by
:
\bea       \label{lowest}
\rho_s(\omega)=-3\left(\frac{e}{4\pi}\right)^2
\left[r\,m\left(1-\frac{m^2}{\omega}\right)+
\int_{m^2}^{\omega} d\alpha \SG_s(\alpha)\left(1-\frac{\alpha}{\omega}\right)\right].
\eea
Note here, that the Pauli-Villars regularization technique
was chosen for  convenience only. It is not so difficult 
to use, for instance, the dimensional  regularization technique. Of course, making direct algebraic subtraction
of unregularized $a,b$ is also possible. All these purely technically different approaches leads to the same results, that is 
the subject what we exactly understand under the  statement  "regularization independence".

When the photon propagator $G_o$ is changed by the introduction of small mass parameter $\lambda$ 

\be
G_o^{\mu\nu}(k)=\frac{-g^{\mu\nu}+k^{\mu}k^{\nu}/k^2}{k^2-\lambda^2+i\ep}
\ee
then the previous results are slightly modified. Repeating the steps as above
it leads to the same form of dispersion relation (\ref{abdisperze}) but with different $\rho_s$,$\rho_v $
, both of them are nonzero now. We get for them

\bea  \label{hmotny}
\rho_s(\omega)&=&- 3\left(\frac{e}{4\pi}\right)^2\left\{mX(\omega,m^2,\lambda^2)+
\int d\alpha \sigma_s(\alpha)X(\omega,\alpha,\lambda^2)\right\}
\nn \\
\rho_v(\omega)&=&\left(\frac{e}{4\pi}\right)^2\Biggl\{X(\omega,m^2,\lambda^2)-2X_1(\omega,m^2,\lambda^2)
\nn \\
&+&(\omega-m^2)\frac{X_1(\omega,m^2,\lambda^2)-2X_1(\omega,m^2,0)}{\lambda^2}
+\int d\alpha \sigma_s(\alpha)\Bigl[X(\omega,\alpha,\lambda^2)
\nn \\
&-&2X_1(\omega,\alpha,\lambda^2)
+(\omega-\alpha)\frac{X_1(\omega,\alpha,\lambda^2)-2X_1(\omega,\alpha,0)}{\lambda^2}\Bigr]\Biggr\}
\eea

Stressed here, that the limit $\lambda\rightarrow 0$ can be safely 
performed an leads to the result (\ref{lowest}) and  $\rho_v= 0$.

\section{Numerical details } 

The unitary equations have been solved by the method of numerical iteration.
To achieve a reasonable accuracy   we should carefully
perform the principal value integrations labeled by  $[\SG,\rho],([\SG,\rho])$ in the equation  for fermion (pseudoscalar) weights.
   For instance the numerical P. integration
\be \label{Delta}
[\SG*\rho] = P. \int\limits_{thresh.}^{\infty}
dx\frac{\SG(s)\rho(x)\frac{s-\mu^2}{x-\mu^2}+\SG(x)\rho(s)}{s-x}.
\ee
is proceed by the following way
which is  based on the exact relation $0=\int\limits_{0}^{\infty}dx\frac{1}{x^2-a^2}$.
Hence we can write for P. integral in (\ref{Delta}):
\bea\label{metallica}
P.\int\limits_c^{\infty}dx\frac{f(x)}{a-x}=
P.\int\limits_c^{\infty}dx\frac{f(x)(a+x)-f(a)2a}{a^2-x^2}+f(a)\ln(\frac{a-c}{a+c})
\eea
where
$$
f(x)=\SG(a)\rho(x)\frac{a-\mu^2}{x-\mu^2}+\SG(x)\rho(a).
$$
The right hand side of the identity (\ref{metallica})
is  particularly useful when evaluated numerically i.e., when $\int \rightarrow \Sigma$.

Using some contemporary PC machine  the  
 criterion $\SG^2_{n,n-1} \simeq 10^{-18}$
can be     achieved, noting that the typical  CPU time  is about $10^2 s$ for a grid number  of   
several hundred mesh points. 
Here, $\SG^2_{n,n-1}$ represents the error between the solution of $n$ times and $n-1$ times iterated unitary equation i.e.,
\be
\SG^2_{n,n-1}=\frac{\int (\rho_n^2-\rho_{n-1}^2)}
{ \int (\rho_n^2+\rho_{n-1}^2)}.
\ee

%&&&&&&&&&&&&&&&&&&&&&&&&&&&&&&&&&&&&&&&&&&&&&&&&&&&&&&&&&&&&&&&&&&&&&&&&&&&&&&&&&&&&&&&&&&&&&&&&&&&&&&&&&&&&&&&&&&&&&&&&&&&&&&&&

%%%%%%%%%%%%%%%%%%%%%%%%%%%%%%%%%%%%%%%%%%%%%%%%%%%%%%%%%%%%%%%%%%%%%%%%%%%%%%%%%%%
%  ______________________appended figures_________________________________________%
%%%%%%%%%%%%%%%%%%%%%%%%%%%%%%%%%%%%%%%%%%%%%%%%%%%%%%%%%%%%%%%%%%%%%%%%%%%%%%%%%%%

\begin{figure}
%\includegraphics{Fig1.eps}
%{{{fig1.eps,angle=270}} }
\centerline{\epsfig{figure=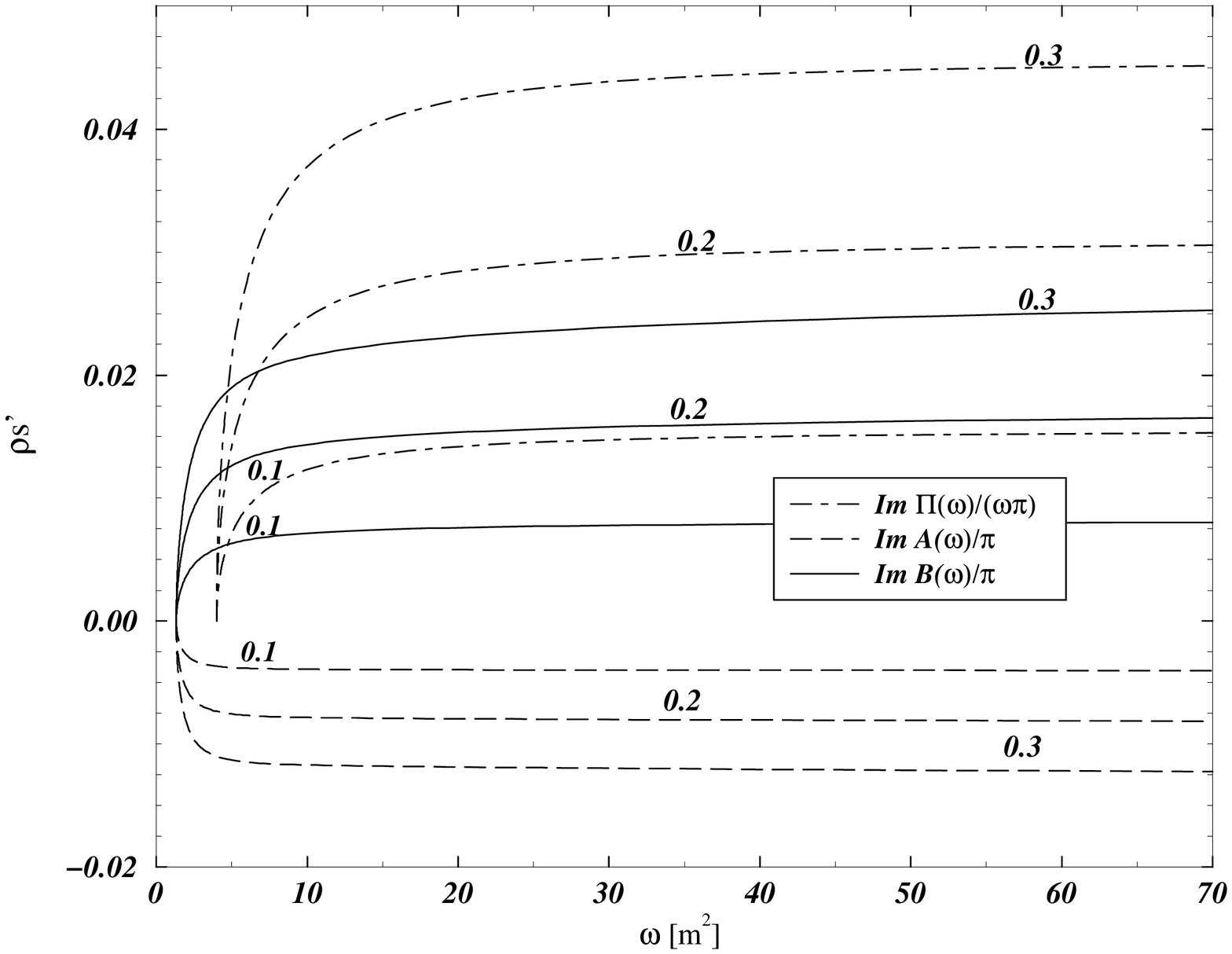,width=8.5truecm,height=12.5truecm,angle=270} }
\caption [caption] { The spectral functions for Yukawa model calculated for the values 
$0.1; 0.2; 0.3 $ of coupling strength
$\lambda=\frac{g^2}{4\pi}$. The lines with the  threshold  $(m+m_{\phi})^2=1.15^2m^2$ correspond with
the Dirac (negative lines) and scalar (positive lines).
The  Lehmann weights for pseudoscalar propagator have the 
appropriate thresholds at $4m^2$ and they are 
always positive.}
\end{figure}
\begin{figure}
\centerline{\epsfig{figure=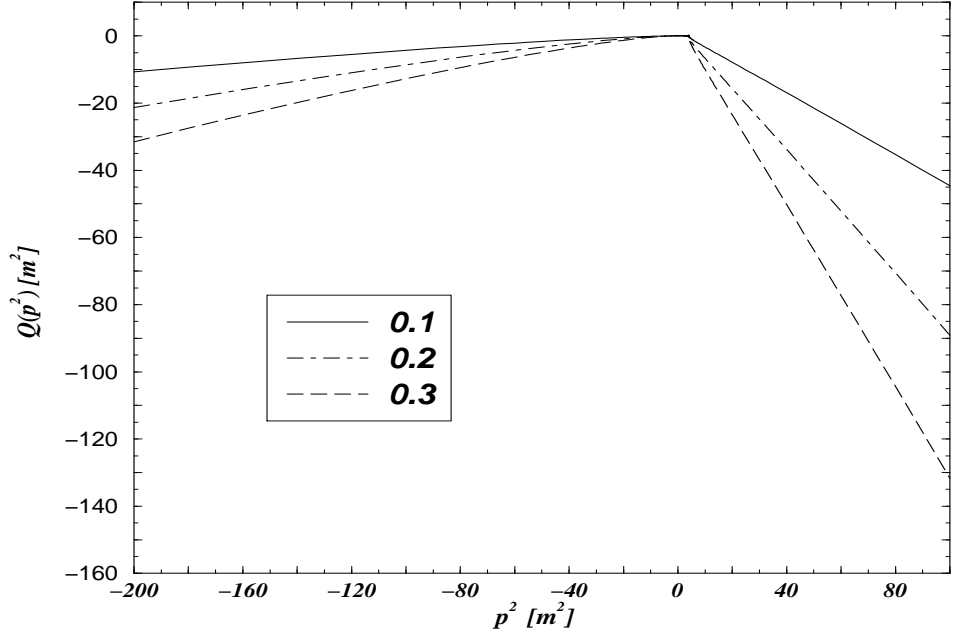,width=8.5truecm,height=12.5truecm,angle=270} }
\caption [caption]{ The self-energy of  pseudoscalar meson 
$Q(p^2)=m^2_{\phi}(\mu)+\Pi_R(\mu^2,p^2)$ for various coupling strengths 
$\lambda$ of Yukawa interaction. The only real part is displayed above the threshold $4m^2$.}
\end{figure}
\newpage
\begin{figure}
\centerline{\epsfig{figure=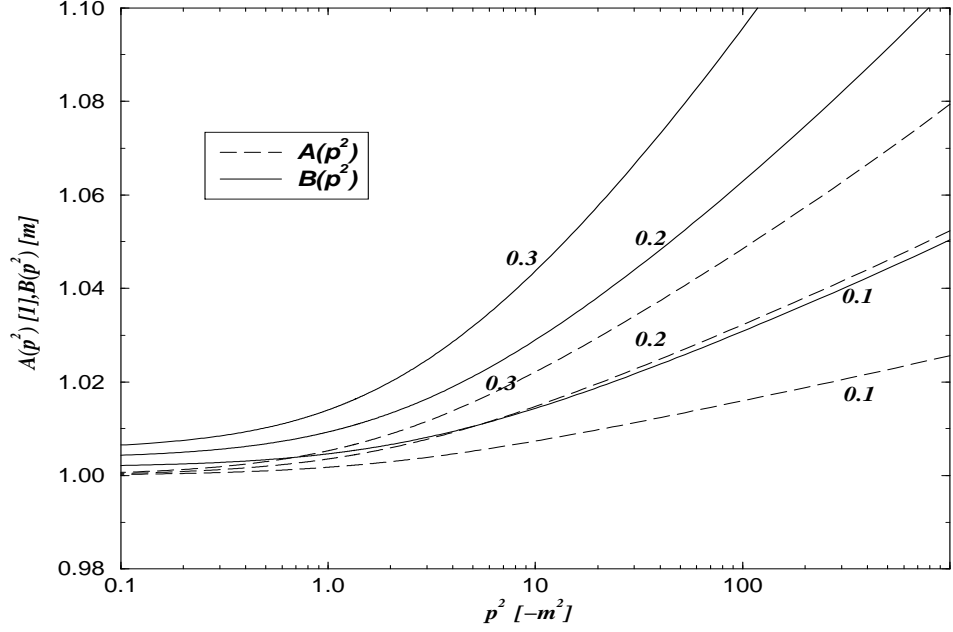,width=8.5truecm,height=12.5truecm,angle=270} }
\caption[caption] {Yukawa fermion functions $A(p^2),B(p^2)$ at  space-like  regime of momentum.}
\end{figure}
\begin{figure}
\centerline{\epsfig{figure=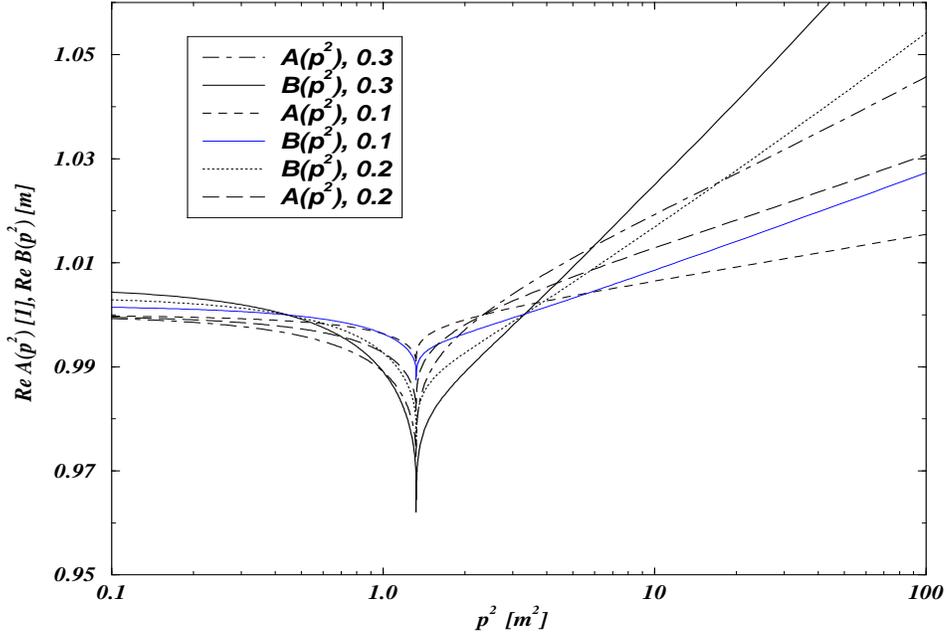,width=8.5truecm,height=12.5truecm,angle=270} }	     
\caption[caption] {Yukawa fermion functions $A(p^2),B(p^2)$ at  time-like  regime of momentum.}
\end{figure}
\begin{figure}
\centerline{\epsfig{figure=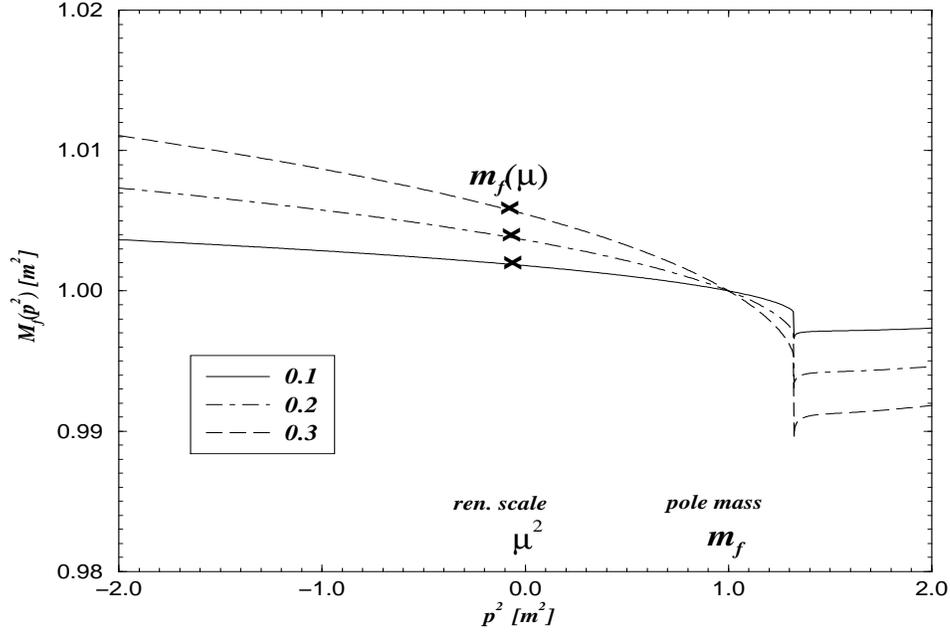,width=8.5truecm,height=12.5truecm,angle=270} }
\caption[caption]{The renormalization scale invariant dynamical mass of Yukawa fermion.
The renormalized mass $m(0)$ is fixed at $\mu=0$ such that the pole mass is 
$m_f=1$ for all the couplings
$\lambda=0.1; 0.2; 0.3 $  }
\end{figure}
\begin{figure}
\centerline{\epsfig{figure=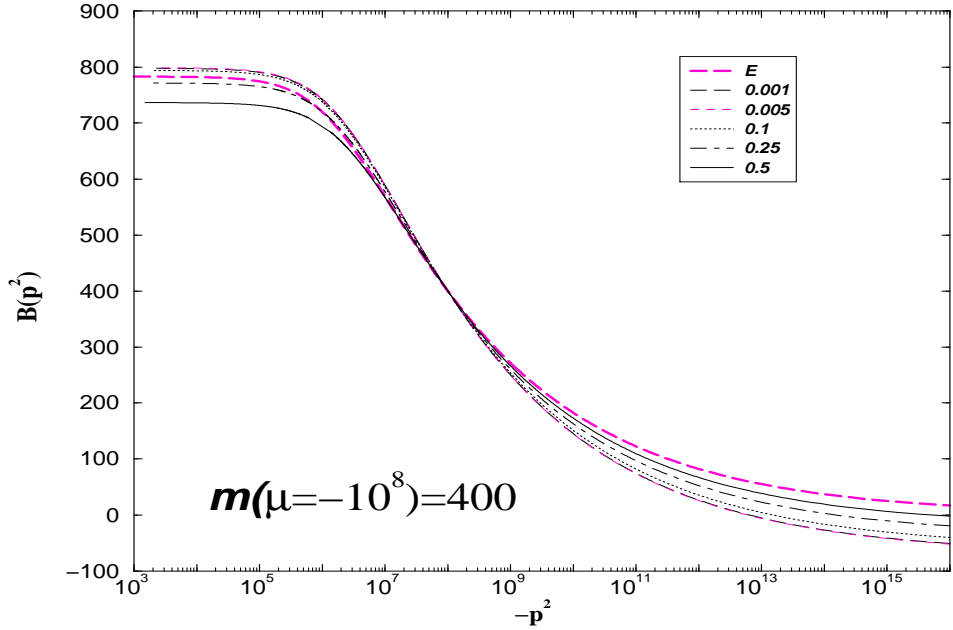,width=8.5truecm,height=12.5truecm,angle=270} }
\caption[caption]{
The dynamical mass of the electron in massive photon QED as they have been 
obtained by solving  modified unitary equtaions.
 The various lines are labeled by the mass of the photon $\lambda$
which is written in the units of an on-shell electron mass. All solutions have the 
coupling value  $\alpha=0.6$, renormalization point
$\mu^2=-10^8$, and renormalized mass is $m(\mu)=400$. The results for two smallest 
$\lambda$ are  not distinguishable. Thin long dashed line labeled by 
the letter E represents the solution obtained in Euclidean formalism. 
In this case, photon was exactly massless.   
}
\end{figure}
\begin{figure}
\centerline{\epsfig{figure=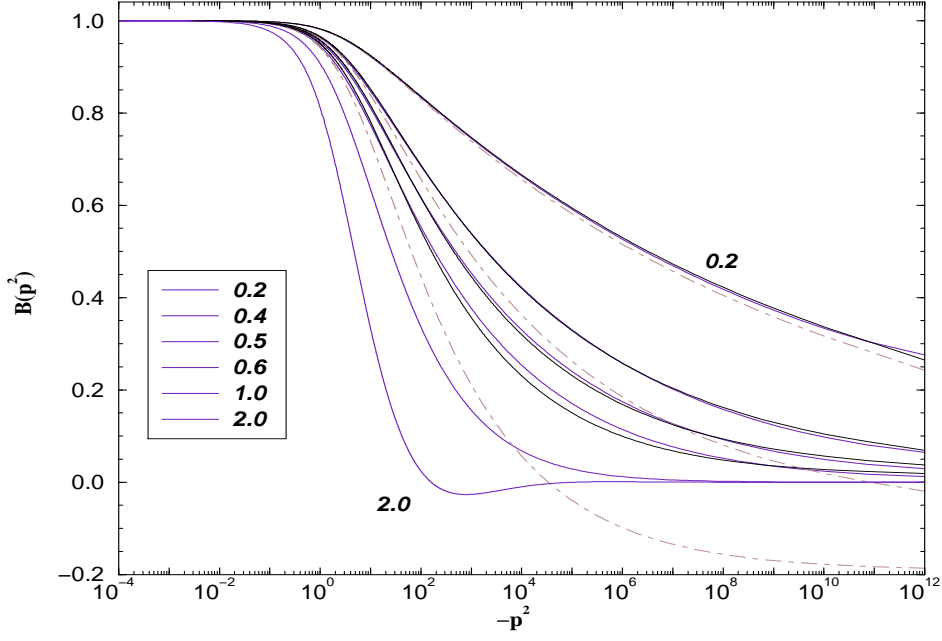,width=8.5truecm,height=12.5truecm,angle=270} }
\caption[caption]{
The electron dynamical mass $M(p^2)$ for space-like  regime of momentum. 
Six  dotted lines represent the Euclidean solutions of momentum DSE.
They are labeled by the value of coupling constant at  given order (from up to down). 
The solutions of unitary equations (dot dashed lines) and the ones
of modified unitary equations (solid thin lines) are added for the coupling $\alpha=0.2;0.4;0.6$.   }
\end{figure}
\begin{figure}
\centerline{\epsfig{figure=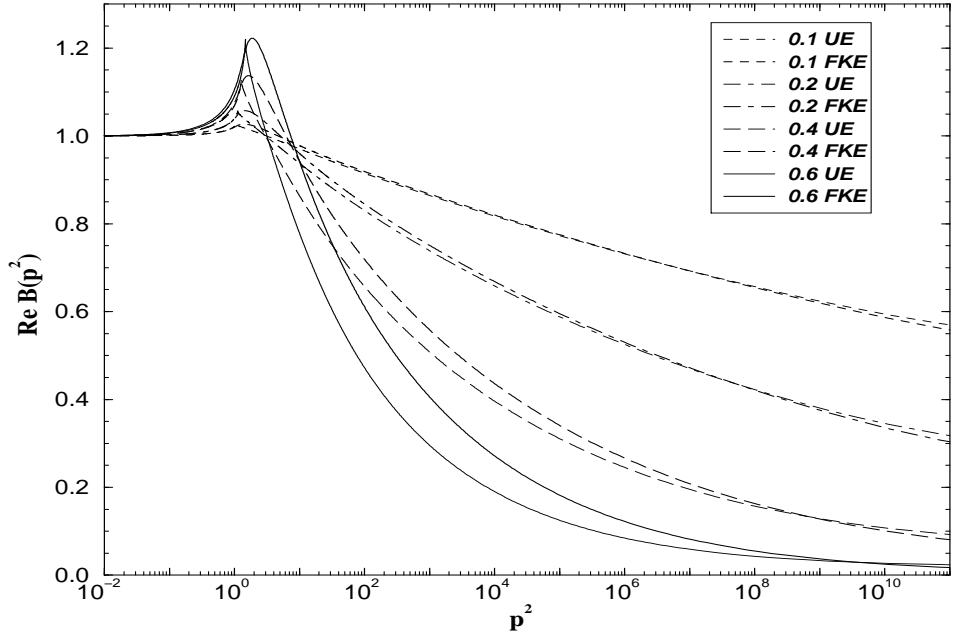,width=8.5truecm,height=12.5truecm,angle=270} }
\caption[caption]{The electron dynamical mass $M(p^2)=B(p^2)$ for 
time-like regime of momentum and renormalization choice $m(0)=1$.
 The thick lines represent Euclidean solutions obtained by principal value 
 integration of the Fukuda-Kugo equation.
 The thin lines correspond with the solutions of modified unitary equations (labeled by UE). }
\end{figure}
\begin{figure}
\centerline{\epsfig{figure=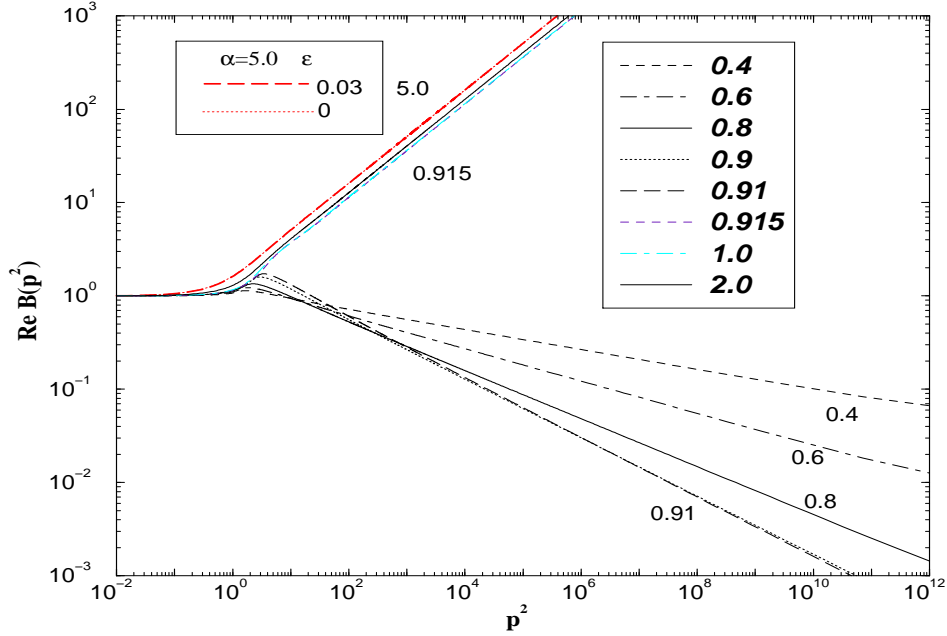,width=8.5truecm,height=12.5truecm,angle=270} }
\caption[caption]{The same as in the previous figure, but for stronger couplings.
The results are obtained from solution of Fukuda-Kugo equation only. 
Signal for confinement is apparent for the coupling $\alpha>0.915$. 
The lines represent a good estimate of true FKE solution for subcritical couplings
but they must be rather exact for the supercritical couplings solutions 
(for the meaning of $\epsilon$ see (\ref{epsik})). }
\end{figure}
\begin{figure}
\centerline{\epsfig{figure=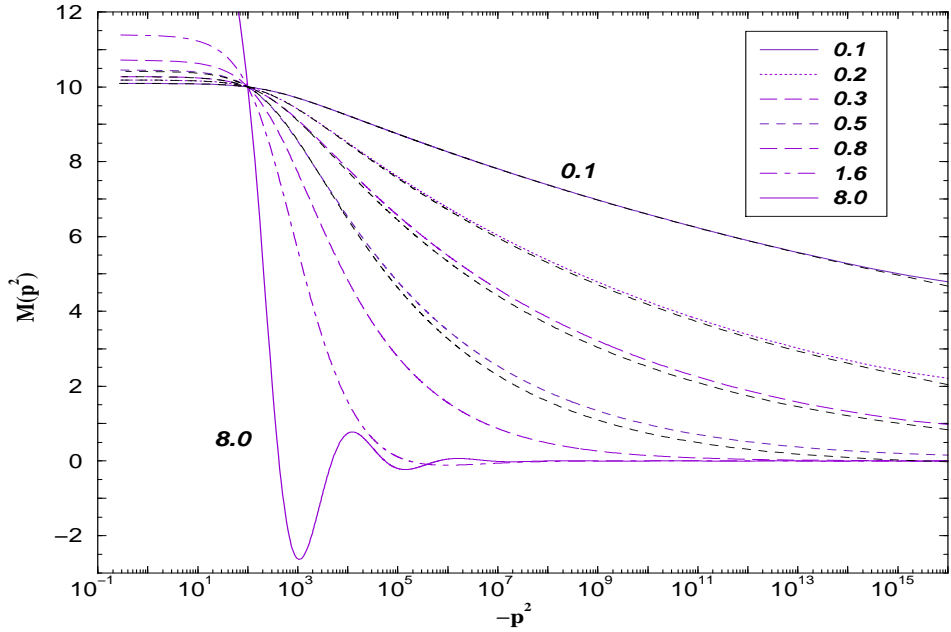,width=8.5truecm,height=12.5truecm,angle=270} }
\caption[caption]{Space-like solutions for electron propagator. 
All solutions have the  renormalization point
$\mu^2=-100$, and renormalized mass  $m(\mu)=10$ and couplings is varied from 
$0.1$ to $8.0$. Note that there are several zero-crossing when the coupling
is large enough. The solutions of modified unitary equations 
(dashed thin lines which  are closed to their Euclidean counter-partners) 
are added for the lowest couplings $\alpha=0.1,0.2,0.3,0.5$.    }
\end{figure}
\begin{figure}
\centerline{\epsfig{figure=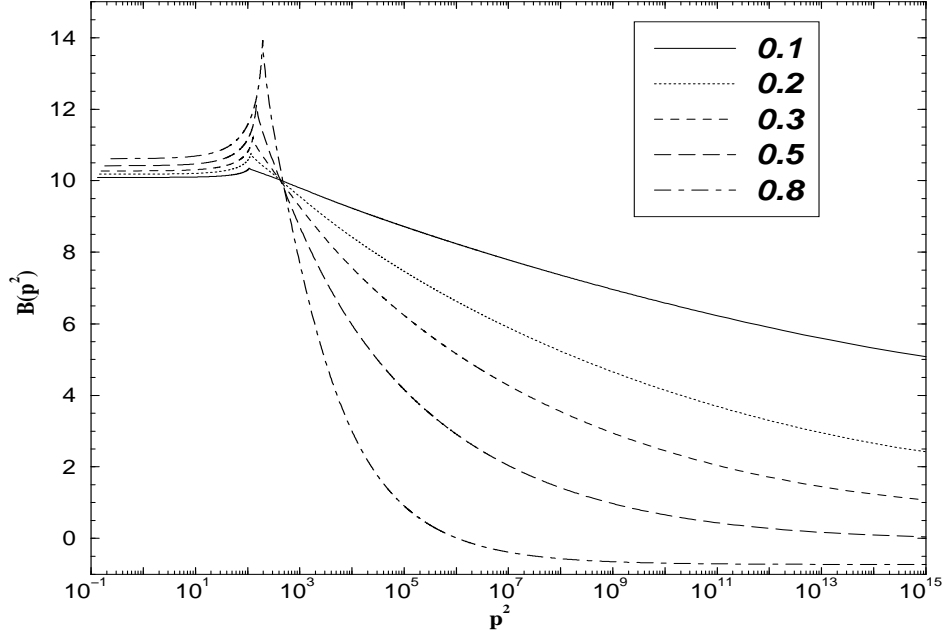,width=8.5truecm,height=12.5truecm,angle=270} }
\caption[caption]{Time-like solution for electron propagator as 
they have been obtained from modified unitary equations.
The renormalization choice is the same as in the previous figure, i.e. $m(-100)=10$.}
\end{figure}
\begin{figure}
\centerline{\epsfig{figure=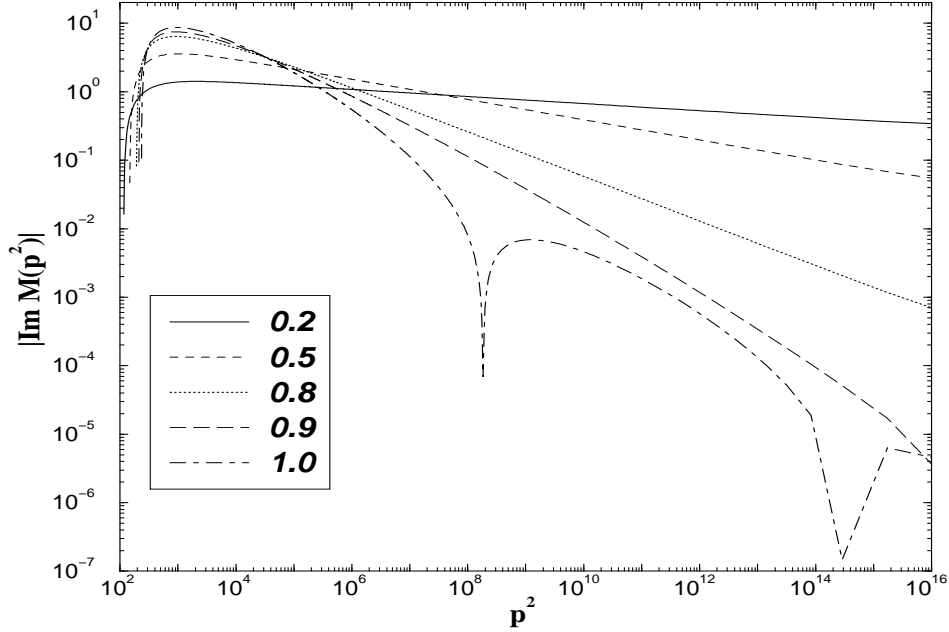,width=8.5truecm,height=12.5truecm,angle=270} }
\caption[caption]{The absolute values of absorptive part of electron self-energy. 
In order to obtain the result for $\alpha=1$, we used the infrared
 cut-off value $0.2m$ in  the modified unitary equations treatment 
 ($m$ is a pole mass, here always $m\simeq m(-100)=10$, for instance $m=14$ for $\alpha=0.8$. }
\end{figure}
\end{document}